\documentclass[aps,prl,showpacs,twocolumn,superscriptaddress,floatfix,tightenlines,amsmath,amssymb,longbibliography,nofootinbib]{revtex4-2}

\usepackage{chemformula} 
\usepackage[T1]{fontenc} 
\usepackage{bibunits}
\defaultbibliography{Refs_verified}  
\defaultbibliographystyle{apsrev4-2}  

\usepackage{graphicx}
\usepackage{color}
\usepackage{dcolumn}
\usepackage{bm}
\usepackage{hyperref}
\usepackage[nolist,nohyperlinks]{acronym}
\usepackage{CJK}
\usepackage{braket}
\usepackage{nicefrac}
\usepackage{amsmath}
\definecolor{cblue}{RGB}{19,107,192}
\hypersetup{colorlinks=true,linkcolor=cblue,citecolor=cblue,urlcolor=cblue}

\newcommand{\MnSn}{Mn$_{3}$Sn}
\newcommand{\MnDSn}{Mn$_{3+\delta}$Sn}

\newcommand{\hexsg}{$P6_{3}/mmc$}

\begin{document}
\raggedbottom
\begin{bibunit}
\title{Anomalous Hall Response Induced by Correlated Disorder \\in the Breathing Kagome Lattice Mn$_{3}$Sn}

\author{Tsung-Han Yang}
\thanks{These authors contributed equally to this work.}
\affiliation{Neutron Scattering Division, Oak Ridge National Laboratory, Oak Ridge, Tennessee 37831, USA}

\author{Seng Huat Lee}
\thanks{These authors contributed equally to this work.}
\affiliation{2D Crystal Consortium, Materials Research Institute, The Pennsylvania State University, University Park, Pennsylvania 16802, USA}
\affiliation{Department of Physics, The Pennsylvania State University, University Park, Pennsylvania 16802, USA}

\author{Hengxin Tan}
\affiliation{Department of Condensed Matter Physics, Weizmann Institute of Science, Rehovot 7610001, Israel}

\author{Yuanpeng Zhang}
\affiliation{Neutron Scattering Division, Oak Ridge National Laboratory, Oak Ridge, Tennessee 37831, USA}

\author{Benjamin A. Frandsen}
\affiliation{Department of Physics and Astronomy, Brigham Young University, Provo, UT 84602, USA}

\author{Václav Petříček}
\affiliation{Institute of Physics of the Czech Academy of Sciences, Prague, Czech Republic}

\author{Huibo Cao}
\affiliation{Neutron Scattering Division, Oak Ridge National Laboratory, Oak Ridge, Tennessee 37831, USA}

\author{Daniel Olds}
\affiliation{National Synchrotron Light Source II, Brookhaven National Laboratory, Upton, New York 11973, USA}

\author{Matthew G. Tucker}
\affiliation{Neutron Scattering Division, Oak Ridge National Laboratory, Oak Ridge, Tennessee 37831, USA}

\author{Jiaqiang Yan}
\affiliation{Materials Science and Technology Division, Oak Ridge National Laboratory, Oak Ridge, Tennessee 37831, USA}

\author{Binghai Yan}
\email{binghai.yan@psu.edu}
\affiliation{Department of Physics, The Pennsylvania State University, University Park, Pennsylvania 16802, USA}

\author{Zhiqiang Mao}
\email{zim1@psu.edu}
\affiliation{2D Crystal Consortium, Materials Research Institute, The Pennsylvania State University, University Park, Pennsylvania 16802, USA}
\affiliation{Department of Physics, The Pennsylvania State University, University Park, Pennsylvania 16802, USA}
\affiliation{Department of Materials Science and Engineering, The Pennsylvania State University, University Park, Pennsylvania 16802, USA}

\author{Qiang Zhang}
\email{zhangq6@ornl.gov}
\affiliation{Neutron Scattering Division, Oak Ridge National Laboratory, Oak Ridge, Tennessee 37831, USA}

\date{\today}

\begin{abstract}
Macroscopic transport tensors are generally constrained by the average crystallographic and magnetic symmetries of a material.
In the kagome antiferromagnetic Weyl semimetals Mn$_{3+\delta}X$ ($X=$~Sn or Ge), previous studies showed that the anomalous Hall conductivity $\sigma_{yx}$ is forbidden by the average \hexsg{} structure and coplanar inverse-triangular magnetic order.
Here we report that nearly stoichiometric Mn$_3$Sn nevertheless exhibits a finite $\sigma_{yx}$ with large hysteresis, together with enhanced $\sigma_{zx}$ and $\sigma_{yz}$, in the inverse-triangular phase below $T_{\mathrm{N1}}\approx 440~\mathrm{K}$, whereas all AHE components vanish in the amplitude-modulated conical phase below $T_{\mathrm{N2}}\approx 280~\mathrm{K}$.
Total scattering and magnetic pair distribution function analysis reveal correlated orthorhombic distortions and noncoplanar Mn moments.
First-principles calculations show that this coupled lattice-spin distortion activates the average symmetry forbidden $\sigma_{yx}$ within the inverse-triangular phase.
Its disappearance below $T_{\mathrm{N2}}$ indicates that the correlated disorder must cooperate with a long-range inverse-triangular antiferromagnetic order capable of supporting Berry curvature.
Our results establish correlated disorder as an active symmetry-breaking degree of freedom that enables topological transport inaccessible from the Bragg-average structure alone.
\end{abstract}

\maketitle
\begingroup
\renewcommand{\thefootnote}{}
\footnotetext{%
Copyright notice: This manuscript has been authored by UT-Battelle, LLC under Contract No.~DE-AC05-00OR22725 with the U.S.~Department of Energy.
The United States Government retains and the publisher, by accepting the article for publication, acknowledges that the United States Government retains a non-exclusive, paid-up, irrevocable, worldwide license to publish or reproduce the published form of this manuscript, or allow others to do so, for United States Government purposes.
The Department of Energy will provide public access to these results of federally sponsored research in accordance with the DOE Public Access Plan
(\url{https://energy.gov/downloads/doe-public-access-plan}).%
}
\endgroup
The interplay between symmetry and topology underlies a wide range of emergent electronic and transport phenomena in quantum materials~\cite{Schnyder2008,Hasan2010,Qi2011,Armitage2018,Watanabe2018,Tang2019,Bernevig2022}. 
When Bloch states acquire a Berry phase constrained by lattice and magnetic symmetries, the resulting Berry curvature acts as a fictitious magnetic field ($H$) in momentum space, giving rise to anomalous Hall and Nernst effects in metals and semimetals~\cite{Fang2003,Haldane2004,Xiao2010,Nagaosa2010,Ikhlas2017}. 
In ferromagnets, the anomalous Hall effect (AHE) generally scales with the net magnetization~\cite{Karplus1954,Onoda2006}, whereas in many conventional collinear antiferromagnets it vanishes because of symmetry-enforced cancellation~\cite{Chen2014,Zelezny2014}. 
Theoretical studies have shown, however, that noncollinear spin textures on frustrated lattices can generate finite Berry curvature even without net magnetization~\cite{Chen2014,Kubler2014}, enabling antiferromagnets to exhibit large anomalous Hall and Nernst responses~\cite{Ikhlas2017,Li2017,Nayak2016}. 
This prediction was realized in the kagome antiferromagnet Mn$_{3+\delta}X$ ($X=$~Sn or Ge), which exhibits a large AHE and anomalous Nernst effect despite a small net moment~\cite{Nakatsuji2015,Kiyohara2016,Ikhlas2017,Wuttke2019,Nayak2016}. 
Angle-resolved photoemission and magnetotransport experiments subsequently established Mn$_{3+\delta}X$ ($X=$~Sn or Ge) as a prototypical magnetic Weyl semimetal, in which Weyl nodes near the Fermi level act as intense monopole sources of Berry curvature~\cite{Kuroda2017,Yang2017}.

Topological electronic properties are often robust against weak perturbations and random disorder, provided that the symmetries protecting the underlying band topology are preserved~\cite{Prodan2011}.
In correlated materials, however, short-range correlations can locally lower symmetry and reorganize coupled lattice, spin, and electronic degrees of freedom without producing a long-range distortion detectable by Bragg diffraction~\cite{Keen2015}.
A macroscopic tensor component forbidden by the average symmetry may therefore become finite when correlated local distortions remove the relevant symmetry constraint over the characteristic length scale of the response.
Mn$_{3}$Sn provides a particularly revealing setting because its Berry-curvature-driven anomalous Hall tensor is tightly constrained by the combined crystal and magnetic symmetries~\cite{Nakatsuji2015,Kuroda2017,Liu2017,Zhang2017}. 

Depending on the deviation $\delta$ from the nominal Mn$_{3}$Sn stoichiometry, Mn$_{3+\delta}$Sn compounds have historically been categorized into A-type and B-type~\cite{Kren1975,Feng2006,Park2018}.
All Mn$_{3+\delta}$Sn compounds first develop the same coplanar inverse-triangular antiferromagnetic order below $T_{\mathrm{N1}}$~\cite{Tomiyoshi1982,Brown1990,Nakatsuji2015,Park2018,Chen2024}.
However, their low-temperature evolution bifurcates: B-type samples ($\delta\approx$0.22-0.33) retain this magnetic order before entering a cluster-glass state near $50~\mathrm{K}$~\cite{Feng2006,Nakatsuji2015,Ikhlas2017}, whereas A-type samples with lower $\delta$ transform into an incommensurate modulated magnetic state below $T_{\mathrm{N2}}\approx280$~K, previously described as a helical phase~\cite{Cable1993,Park2018,Chen2024}. In the inverse-triangular antiferromagnetic phase, symmetry permits a finite $\sigma_{zx}$~\cite{Zhang2017}, whereas $\sigma_{yx}$ is forbidden. Previous measurements on B-type samples~\cite{Nakatsuji2015,Liu2017} revealed finite $\sigma_{zx}$ and  $\sigma_{yz}$ with narrow hysteresis, as well as a weak $H$-linear $\sigma_{yx}$. Note that this linear $\sigma_{yx}$ is purely field-induced and the spontaneous zero-field $\sigma_{yx}$ indeed remains zero.
In contrast, the nearly stoichiometric \(\mathrm{Mn_3Sn}\) crystals studied here exhibit a  $H$-nonlinear  $\sigma_{yx}$ with hysteresis, alongside enhanced $\sigma_{zx}$ and $\sigma_{yz}$ that display similar narrow hysteresis, despite the absence of average structural symmetry breaking.
We demonstrate that these three tensor components have distinct origins.
By resolving the crystal and magnetic structures across long-range and local length scales, we reveal correlated local orthorhombic distortions and Mn-spin canting, likely associated with a pronounced breathing distortion of the kagome lattice, which lift the symmetry constraints imposed by the average structure and activate the nominally forbidden Hall channel.

\begin{figure}[ht!]
    \includegraphics[width=1\linewidth]{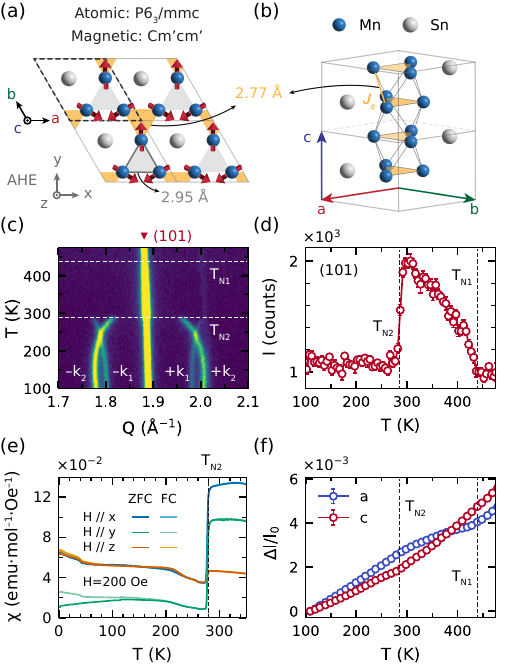}
\caption{Crystal and magnetic structures from diffraction and magnetization measurements.
(a) and (b) show top and side views, respectively at 295~K.
(c) Temperature-dependent neutron powder diffraction at POWGEN showing the emergence of incommensurate magnetic reflections below $T_{\mathrm{N2}}$.
(d) Temperature dependence of the integrated intensity of the (101) reflection, revealing the magnetic phase transitions at $T_{\mathrm{N1}}$ and $T_{\mathrm{N2}}$.
(e) Temperature dependence of the magnetization for $H\parallel x$, $y$, and $z$, showing the magnetic transition at $T_{\mathrm{N2}}$.
(f) Temperature dependence of the lattice parameters obtained from synchrotron powder x-ray diffraction performed at 28-ID-1 (PDF) beamline.
}
\label{fig:AvgStrs}
\end{figure}

We first investigate the chemical composition, structure and magnetic transitions of our sample.
Wavelength-dispersive x-ray spectroscopy (WDS) measurements yield a composition of Mn$_{3.002}$Sn, as summarized in Table~S1 of the Supplemental Material (SM)~\cite{SMat}, consistent with the composition refined from single-crystal x-ray and neutron diffraction.
The single-crystal x-ray diffraction data at $295~\mathrm{K}$ ($T_{\mathrm{N2}}<T<T_{\mathrm{N1}}$), and $220~\mathrm{K}$ ($T<T_{\mathrm{N2}}$) are well described by the previously reported \hexsg{} structure~\cite{Park2018,Chen2024} [Fig.~\ref{fig:AvgStrs}(a) and Tables S2 and S3 of the SM~\cite{SMat}].
Notably, the refined structure at $295~\mathrm{K}$ exhibits a pronounced in-plane Mn--Mn bond disproportionation, with bond lengths of $2.77$ and $2.95$~\AA{}, corresponding to strongly contracted and expanded Mn triangles within the breathing kagome lattice [Fig.~\ref{fig:AvgStrs}(a)].
This large breathing amplitude contrasts with the substantially weaker bond disproportionation reported in B-type samples~\cite{Nakatsuji2015,Ikhlas2017,Park2018}, despite the preservation of the same average \hexsg{} symmetry.

Neutron powder diffraction measurements at POWGEN show that below $T_{\mathrm{N1}}\approx440~\mathrm{K}$, $\mathbf{k}=0$ magnetic Bragg scattering develops as shown in Fig.~\ref{fig:AvgStrs}(c--d).
The coplanar inverse-triangular magnetic order is shown in Fig.~\ref{fig:AvgStrs}(a) and is known to host a small in-plane net ferromagnetic component~\cite{Nakatsuji2015,Ikhlas2017,Park2018}.
Upon cooling below $T_{\mathrm{N2}} \approx 280~\mathrm{K}$, the $\mathbf{k}=0$ magnetic contribution to the nuclear reflections disappears and is replaced by incommensurate magnetic satellites indexed by two propagation vectors, $\mathbf{k}_1$ and $\mathbf{k}_2$, as well as a higher-order harmonic at $2\mathbf{k}_1 + \mathbf{k}_2$ [Fig.~\ref{fig:AvgStrs}(c) and Fig.~\ref{fig:EM:TN2}(a) in the End Matter].
The $\mathbf{k}_1$ and $\mathbf{k}_2$ components exhibit opposite temperature dependences and cross near $\approx 240~\mathrm{K}$ before settling at $\mathbf{k}_1 = (0,~0,~0.0878)$ and $\mathbf{k}_2 = (0,~0,~0.1049)$ at $5~\mathrm{K}$.
The refined magnetic structure is shown in Fig.~\ref{fig:EM:TN2}(c) in the End Matter.
This amplitude-modulated conical structure consists of an intermodulation of an out-of-plane longitudinal spin density wave with $\mathbf{k}_1$, an in-plane elliptical helix with $\mathbf{k}_2$, and a second in-plane elliptical helix arising from $2\mathbf{k}_1 + \mathbf{k}_2$ [Fig.~\ref{fig:EM:TN2}(d--f) in the End Matter and Table S4 in the SM~\cite{SMat}].
Further details of the magnetic structures above and below $T_{\mathrm{N2}}$, together with comparisons with previously reported models, are provided in the End Matter.

At $T_{\mathrm{N2}}$, magnetization measured for $H\parallel x$, $y$, and $z$ exhibits a clear anomaly [Fig.~\ref{fig:AvgStrs}(e)].
Throughout both transitions, the average crystallographic symmetry remains \hexsg{}.
Nevertheless, synchrotron x-ray powder diffraction reveals distinct anomalies in the in-plane lattice parameter $a(=b)$ at both $T_{\mathrm{N1}}$ and $T_{\mathrm{N2}}$, while the out-of-plane lattice parameter $c$ exhibits a change in slope at $T_{\mathrm{N2}}$ [Fig.~\ref{fig:AvgStrs}(f)].
These lattice anomalies demonstrate the long-range coupling between spin and lattice degrees of freedom.

\begin{figure}[ht!]
    \includegraphics[width=1\linewidth]{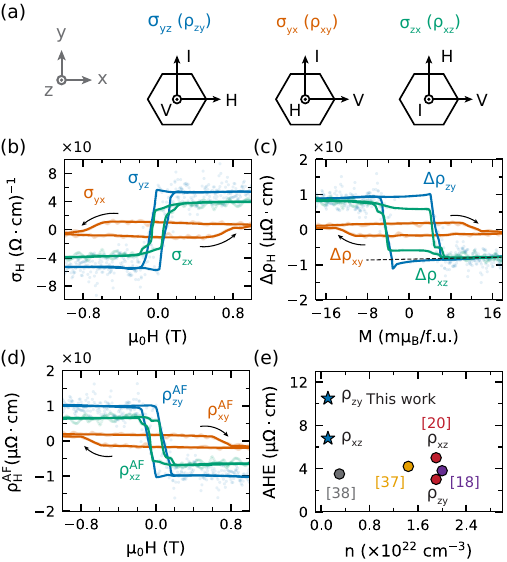}
\caption{Magnetic field dependence of the AHE in nearly stoichiometric Mn$_{3}$Sn at 300~K. 
(a) Schematics of the three configurations used for AHE measurements on    $\sigma_{yz}$, $\sigma_{yx}$ and $\sigma_{zx}$.
(b) Field-dependent Hall conductivity components measured at $300~\mathrm{K}$.
(c) Anomalous Hall resistivity $\Delta\rho_{H}=\rho_{H}-R_{0}\mu_{0}H$ plotted as a function of magnetization $M$.
(d) Field dependence of the antiferromagnetic Hall component, $\rho_{H}^{\mathrm{AF}}=\rho_{H}-R_{0}\mu_{0}H-R_{s}\mu_{0}M$, obtained by subtracting the ordinary and magnetization-linear contributions.
(e) Comparison of the anomalous Hall resistivity and carrier density $n$ for {\MnDSn} reported in this work and previous studies.
}
\label{fig:AHE}
\end{figure}

Hall measurements across the magnetic transitions show that the anomalous Hall response of Mn$_{3}$Sn is confined to the inverse-triangular phase, $T_{\mathrm{N2}}<T<T_{\mathrm{N1}}$, and vanishes upon entering the amplitude-modulated conical magnetic phase below $T_{\mathrm{N2}}$ [Fig.~\ref{fig:EM:MH}(d--f) in the End Matter].
At $300~\mathrm{K}$, we found a sizable $\sigma_{zx}\approx40~\Omega^{-1}\cdot\mathrm{cm}^{-1}$ [Fig.~\ref{fig:AHE}(b)], nearly twice the previously reported values~\cite{Nakatsuji2015,Ikhlas2017,Li2017,Kuroda2017}, together with a finite $\sigma_{yz}$ consistent with earlier measurements on B-type sample~\cite{Nakatsuji2015}.
Both $\sigma_{zx}$ and $\sigma_{yz}$ components exhibit narrow hysteresis, concomitant with reversal of the weak in-plane ferromagnetic moment [Fig.~\ref{fig:EM:MH}(a--b) in the End Matter]. 
Notably, we observe a $H$-nonlinear $\sigma_{yx}\approx12~\Omega^{-1}\cdot\mathrm{cm}^{-1}$, approximately $20\%$ of $\sigma_{zx}$ or $\sigma_{yz}$ [Fig.~\ref{fig:AHE}(b)].
Angle- and temperature-dependent measurements confirm that $\sigma_{yx}$ with large hysteresis is an intrinsic response rather than an artifact (see Fig.~S2 and SM~\cite{SMat} for details).
It is also worth noting that our nearly stoichiometric Mn$_{3}$Sn crystal exhibits the largest reported $\rho_{H}$ and the lowest carrier density among bulk crystals~\cite{Nakatsuji2015,Li2017,Li2023,Yano2024}, as summarized in Fig.~\ref{fig:AHE}(e).

After subtraction of the ordinary Hall contribution, none of the three components of $\Delta \rho_H$ scales proportionally with the magnetization [Fig.~\ref{fig:AHE}(c)].
Upon further subtracting the magnetization-linear contributions, the residual signal $\rho_{\mathrm{H}}^{\mathrm{AF}}$ remains substantial [Fig.~\ref{fig:AHE}(d)]. The hysteretic $\rho_{xy}^{\mathrm{AF}}$ observed here is distinct from the zero $\rho_{xy}^{\mathrm{AF}}$ signal reported previously\cite{Nakatsuji2015,Li2023}. Furthermore, $\sigma_{yx}$ (or $\rho_{xy}^{\mathrm{AF}}$) exhibits a coercive field of approximately $0.8~\mathrm{T}$, substantially larger than those associated with $\sigma_{zx}$ and $\sigma_{yz}$.
In addition, magnetization measured in the same $H\parallel~z$ geometry is nonhysteretic [Fig.~\ref{fig:EM:MH}(c) of End Matter], in contrast to the clear magnetic hysteresis loops observed in the other two configurations.
All these results indicate a distinct origin of $\sigma_{yx}$ from that of $\sigma_{yz}$ or $\sigma_{zx}$.

\begin{figure}[t!]
    \includegraphics[width=1\linewidth]{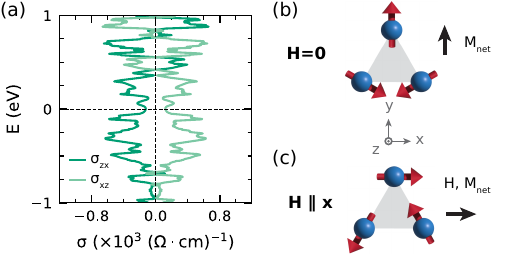}
\caption{
Anomalous Hall conductivity and magnetic-field-induced spin rotation of Mn$_{3}$Sn.
(a) DFT-calculated intrinsic AHC components for the average hexagonal structure.
(b) Reference magnetic structure used for the DFT calculations of the intrinsic AHC.
(c) Schematic rotation of the noncollinear Mn moments under $H\parallel\mathbf{x}$, generating $\sigma_{yz}$.
}
\label{fig:DFT}
\end{figure}

To explore the origins of the three AHE tensor components, we performed DFT calculations based on the average hexagonal structure with the inverse-triangular magnetic order shown in Fig.~\ref{fig:AvgStrs}(a). We confirmed that only $\sigma_{zx}$ is expected to be nonzero, whereas both $\sigma_{yz}$ and $\sigma_{yx}$ are forbidden [Fig.~\ref{fig:DFT}(a)], consistent with previous report~\cite{Zhang2017}. This is because the net in-plane moment $\boldsymbol{M}_{\mathrm{net}} \parallel y$ preserves the vertical glide mirror symmetry $M_y$, which uniquely allows a Berry curvature along $y$ ($\Omega_y \neq 0$) and generates the primary response $\sigma_{zx}$ ($\sigma_{xz} = -\sigma_{zx}$) (see Fig.~\ref{fig:DFT}(b)). However, when a magnetic field is applied along the $x$-axis to measure $\sigma_{yz}$, this magnetic field polarizes the net FM component by rotating the spin configuration by $90^\circ$, aligning $\boldsymbol{M}_{\mathrm{net}} \parallel x$, shifting the invariant mirror plane to $M_x$ ($\Omega_x \neq 0$), and selectively activating $\sigma_{yz}$ ($\sigma_{zy} = -\sigma_{yz}$), as illustrated in  Fig.~\ref{fig:DFT}(c). This may interpret why the coercive fields of  $\sigma_{yz}$ and $\sigma_{zx}$ are nearly identical. Compared with previous reports~\cite{Nakatsuji2015,Li2017,Li2023,Yano2024}, the larger magnitudes of $\sigma_{zx}$ and $\sigma_{yz}$ found here are attributed to changes in the Berry curvature associated with a shift in chemical potential due to the lower Mn content of our crystal compared with B-type crystals.
Nevertheless, in the coplanar spin structures shown in  Fig.~\ref{fig:DFT}(b) or (c), the combined symmetry of $M_z$ and time reversal $\mathcal{T}$ forces  $\sigma_{yx} = 0$. In addition, the $H$-induced spin canting toward the $c$ axis can generate only an $H$-linear $\sigma_{yx}$ as established by both experimental and theoretical studies~\cite{Li2023}. Therefore, the remaining $\sigma_{yx}$ with hysteresis cannot be interpreted within the framework of the average crystal and magnetic structures, which motivates us to investigate the local structural and magnetic configurations.

\begin{figure}[ht!]
    \includegraphics[width=1\linewidth]{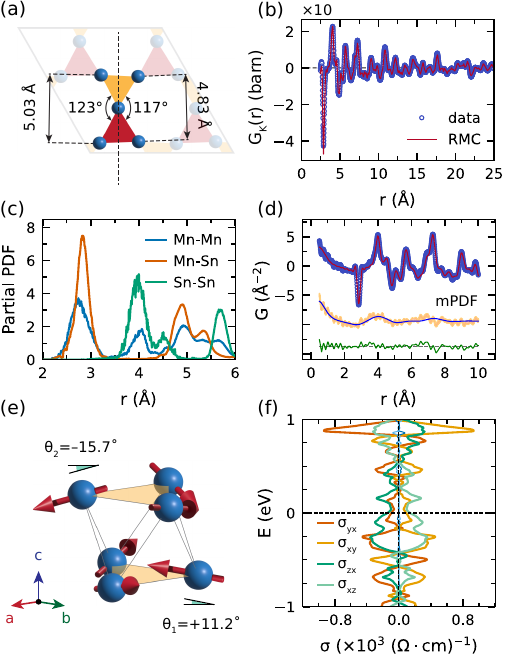}
\caption{Local structural distortion and spin canting in Mn$_{3}$Sn. 
(a) Local structural motif of Mn$_{3}$Sn at 500~K presented in a hexagonal cell. The tilted Mn--Mn--Mn triangles within the kagome plane are resolved, revealing a bond-length disproportionation of 0.2~\AA{} and asymmetric bond angles of 123$^{\circ}$ and 117$^{\circ}$, which breaks hexagonal symmetry. 
(b) RMC modeling results of $G_{\mathrm{K}}(r)$ compared to experimental data. 
(c) Partial pair distribution function for individual atomic pairs, showing well-resolved Mn--Mn and Mn--Sn atomic pairs near 2.7~\AA{}. 
(d) Magnetic PDF fits showing a substantial mPDF signal. 
(e) Local magnetic-moment motif at 300 K from mPDF refinement.
(f) DFT-calculated intrinsic AHC components for the distorted orthorhombic structure with finite out-of-plane canting of the Mn moments.
}
\label{fig:PDF}
\end{figure}

To determine possible short-range lattice distortions, we performed temperature-dependent neutron total-scattering measurements.
We first analyze the 500~K data using reverse Monte Carlo (RMC) modeling with \textsf{RMCProfile}~\cite{Tucker2007,Zhang2020}, well above \(T_{\mathrm{N1}}\), thereby isolating the lattice degree of freedom in the paramagnetic state.
The folded RMC configuration in Fig.~\ref{fig:PDF}(a) reveals locally tilted Mn triangular motifs, producing inequivalent Mn--Mn--Mn bond angles and Mn--Mn distances of 4.83~\AA{} and 5.03~\AA{}.
The converged configuration reproduces both the measured F$_{\mathrm{K}}(Q)$ [Fig.~S4 in the SM] and G$_{\mathrm{K}}(r)$ [Fig.~\ref{fig:PDF}(b)], showing consistency with the reciprocal- and real-space scattering data. 
The partial pair distribution functions in Fig.~\ref{fig:PDF}(c) further resolve the nearest-neighbor Mn--Mn and Mn--Sn correlations as distinct contributions, showing that the local distortions are physically constrained rather than artifacts arising from overlap between neighboring pair correlations.
Symmetry analysis using \textsf{FINDSYM}~\cite{Stokes2005} identifies the dominant local environment as orthorhombic \(Ama2\) (No.~40). Thus, local symmetry breaking is already present in the paramagnetic lattice and persists down to temperatures below $T_{\mathrm{N2}}$ (see Fig.~S5 in the SM~\cite{SMat}).  

To resolve the local magnetic structure, we analyze the magnetic pair distribution function data at 300 K, reduced with \textsf{pyFAI}~\cite{Kieffer2013} and \textsf{PDFgetX3}~\cite{Juhas2013}, using \textsf{diffpy.mpdf}~\cite{Frandsen2022}.
The residual mPDF, obtained by subtracting the fitted atomic PDF from the total PDF, is modeled starting from the inverse-triangular magnetic structure.
A coplanar model, including local in-plane canting, does not adequately reproduce the mPDF data, whereas allowing layer-dependent out-of-plane canting substantially improves the fit.
The fitted mPDF is shown in Fig.~\ref{fig:PDF}(d).
The least-squares fit yields a locally ordered Mn moment of $2.19 \pm 0.08~\mu_{\mathrm{B}}$ and a magnetic correlation length of $\xi = 8.8 \pm 1.7$~\AA{}, indicating that the local noncoplanar correlations extend over only several Mn--Mn spacings while the underlying inverse-triangular order remains long-ranged.
The refinement further gives opposite out-of-plane canting angles of $-15.7 \pm 4.8^\circ$ and $11.2 \pm 6.7^\circ$ [Fig.~\ref{fig:PDF}(e)], revealing a locally noncoplanar magnetic configuration.
A Bayesian Markov-chain Monte Carlo analysis further supports the finite out-of-plane canting, yielding posterior distributions consistent with the least-squares solution and excluding a coplanar configuration at the 95\% credible level [Fig.~\ref{fig:EM:mPDF_Posterior} of End Matter].
The local orthorhombic distortions and noncoplanar Mn moments provide evidence for correlated disorder in our Mn$_{3+\delta}$Sn, given the negligible site disorder for $\delta\approx0$.



Our DFT calculations show that incorporating the orthorhombic distortion and experimentally determined out-of-plane canting produces a finite $\sigma_{yx}$ [Fig.~\ref{fig:PDF}(f)]. This noncoplanar spin arrangement with unequal $c$-axis canting angles breaks the combined symmetry of $M_z$ and time reversal $\mathcal{T}$, activating $\sigma_{yx}$.
Although the average \hexsg{} structure and coplanar inverse-triangular order retain a mirror symmetry that requires $\sigma_{yx}=0$, the locally orthorhombic lattice and noncoplanar spin configuration lift this constraint.
Berry-curvature transport is thus governed not solely by the average crystallographic and magnetic symmetries, but also by the correlated local magnetostructure.
Whereas Bragg diffraction resolves long-range periodic order, electronic transport remains sensitive to local symmetry breaking over the finite length scales relevant to carrier propagation.
Correlated lattice and spin disorder can therefore activate tensor components forbidden by the average symmetry without condensing into a long-range structural phase.
Furthermore, the disappearance of $\sigma_{yx}$ below $T_{\mathrm{N2}}$ indicates that local symmetry lowering alone is insufficient and must act cooperatively with the long-range magnetic order that supports the Berry-curvature response.

The pronounced breathing character of the average kagome lattice provides a possible microscopic origin for the local magnetic noncoplanarity.
As the disparity between contracted and expanded Mn triangles increases, the interlayer Mn--Mn pathway for $J_c$ becomes relatively short compared with the in-plane inter-triangle separation, as shown in Fig.~\ref{fig:AvgStrs}(a) and (b).
This altered hierarchy suggests enhanced interlayer magnetic coupling, which, combined with spin-orbit anisotropic interactions, increases the susceptibility to out-of-plane spin fluctuations and stabilizes locally canted configurations. 
This exchange hierarchy may also underlie the pronounced sensitivity of Mn$_{3}$Sn to excess-Mn concentration, where stoichiometry reorganizes local lattice distortions, magnetic coupling, and spin arrangements without changing the Bragg-average symmetry~\cite{Park2018}.
Consequently, the composition controls topological transport not only by shifting the carrier concentration or chemical potential, but also by tuning the local structural and magnetic environment.


In conclusion, we report $H$-nonlinear and hysteretic  $\sigma_{yx}$, forbidden by average symmetry, alongside the enhanced  $\sigma_{zx}$ and $\sigma_{yz}$ responses, in $T_{\mathrm{N2}} < T < T_{\mathrm{N1}}$ of nearly stoichiometric Mn$_{3}$Sn.
We show that correlated local orthorhombic distortions and noncoplanar Mn moments lift the symmetry constraint on $\sigma_{yx}$ while preserving the allowed $\sigma_{zx}$ and $\sigma_{yz}$ responses.
The disappearance of AHE upon entering the noncoplanar amplitude-modulated conical phase below $T_{\mathrm{N2}}$ further indicates that this forbidden Hall response arises from the cooperative interplay between local symmetry breaking and long-range inverse-triangular order.
Our results establish correlated disorder as an active degree of freedom for reshaping Berry-curvature-driven transport without requiring a long-range crystallographic symmetry-breaking transition, and may also provide a new insight into the puzzling macroscopic anomalous hall effect.  

\section{Data availability}
The datasets generated and analyzed during this study are available in the Zenodo repository at [URL place holder]. 
The repository contains neutron diffraction, total scattering, magnetization, and transport data, as well as density functional theory (DFT) calculation results associated with this work. Source data underlying the figures are provided with this paper.

\section{Acknowledgments}
We thank Xiaoping Wang and Sylwia Pawledzio for their assistance with the benchtop x-ray diffraction measurements at the TOPAZ beamline.
This research used resources at the Spallation Neutron Source (SNS) and the High Flux Isotope Reactor (HFIR), both DOE Office of Science User Facilities operated by Oak Ridge National Laboratory.
Beam time at POWGEN was allocated under proposal numbers IPTS-29547.1 and IPTS-34790.1, and beam time at HB-3A (DEMAND) was allocated under proposal number IPTS-32272.1. J.Y. was supported by the U.S. Department of Energy, Office of Science, Basic Energy
Sciences, Materials Sciences and Engineering Division. 
This research used resources at the 28-ID-1 (PDF) beamline of the National Synchrotron Light Source II, a U.S. Department of Energy (DOE) Office of Science User Facility operated for the DOE Office of Science by Brookhaven National Laboratory under Contract DE-SC0012704.
B. F. was supported by the U.S. Department of Energy, Office of
Science, Basic Energy Sciences (DOE-BES) through Award No. DESC0021134.

\putbib

\clearpage

\onecolumngrid

\section*{End Matter}

\twocolumngrid

\begin{figure}[!t]
    \includegraphics[width=1\linewidth]{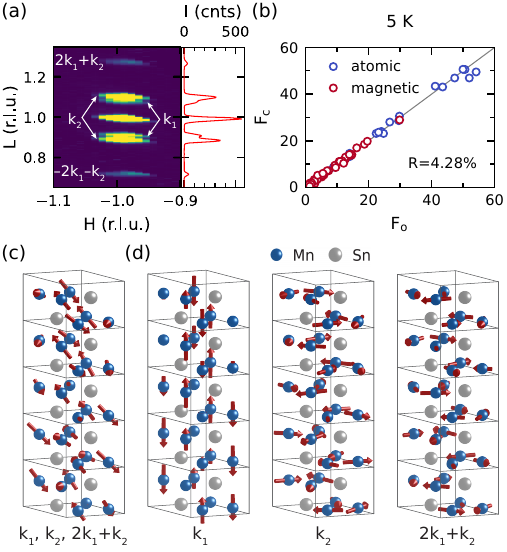}
\caption{
Single-crystal neutron diffraction and low-T magnetic structure of Mn$_3$Sn below $T_{\mathrm{N2}}$.
(a) Reconstruction of the $(H0L)$ plane near the $(-101)$ nuclear reflection at $6~\mathrm{K}$, showing magnetic satellites with $\mathbf{k}_1$ and $\mathbf{k}_2$ and their higher harmonics; the corresponding $L$ profile is shown at right.
(b) Observed and calculated structure-factor amplitudes for nuclear (blue) and magnetic (red) reflections.
(c) Refined magnetic structure combining the $\mathbf{k}_1$, $\mathbf{k}_2$, and $2\mathbf{k}_1+\mathbf{k}_2$ components.
(d) Decomposition into the individual $\mathbf{k}_1$, $\mathbf{k}_2$, and  components.
}
\label{fig:EM:TN2}
\end{figure}
\begin{figure*}[!ht]
    \includegraphics[width=\linewidth]{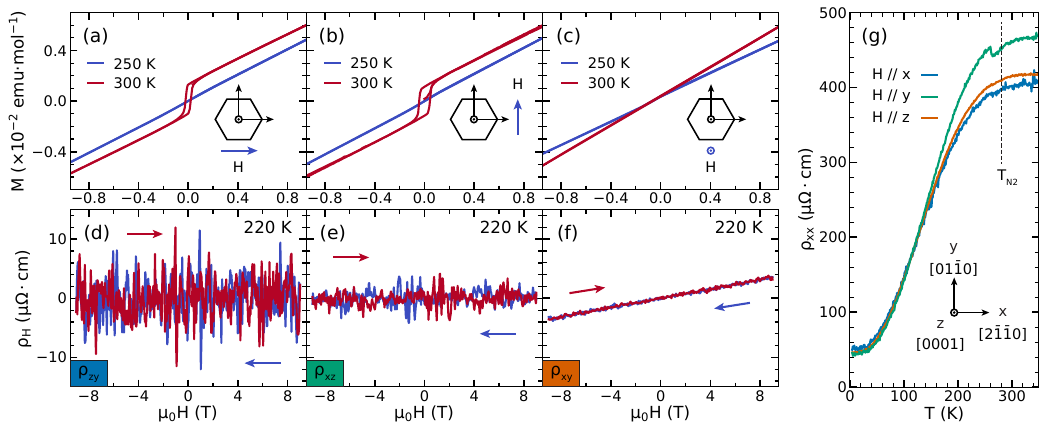}
\caption{
Anisotropic magnetic and Hall responses and temperature-dependent longitudinal resistivity of Mn$_{3}$Sn.
(a)--(c) Field-dependent magnetization $M(H)$ measured at $250$ and $300~\mathrm{K}$ for $H\parallel x$, $H\parallel y$, and $H\parallel z$, respectively.
(d)--(f) Hall resistivity $\rho_H(H)$ measured at $220~\mathrm{K}$, below $T_{\mathrm{N2}}$, for the corresponding field orientations.
Arrows indicate the field-sweep directions.
(g) Temperature-dependent longitudinal resistivity $\rho_{xx}(T)$ measured for $H\parallel x$, $H\parallel y$, and $H\parallel z$.
The dashed line marks $T_{\mathrm{N2}}$.
}
\label{fig:EM:MH}
\end{figure*}

\textit{Magnetic structural determination.}
The Rietveld refinement of the neutron powder diffraction data at 300~K using the magnetic space group $Cm'cm'$ (BNS No.~63.464), is shown in Fig.~S3 of the SM~\cite{SMat}.
The ordered moment on each Mn site is refined to 2.32(2) $\mu_{B}$.
A net FM component observed in the $ab$ plane (see Fig.~\ref{fig:EM:MH}) indicates slight in-plane canting within the nearly 120$^\circ$ triangular AFM structure, consistent with prior diffraction~\cite{Brown1990,Song2020} and DFT results~\cite{Zhang2017}.
An alternative triangular magnetic order with magnetic space group $Cmc'm'$ (BNS No.~63.463) has recently been proposed~\cite{Chen2024,Cederholm2026} (see Fig.~S3(c) in the SM~\cite{SMat}).
This structure can be obtained by rotating the individual spins in the $Cm'cm'$ structure by 90$^\circ$ within the $ab$ plane.
The $Cmc'm'$ model yields a comparable fit quality, and therefore the two models cannot be distinguished in the present work.
However, these two magnetic structures could be reconciled by spin rotation behavior arising from field-induced polarization of the net FM component during polarized neutron experiments ~\cite{Brown1990,Chen2024,Cederholm2026}.
We adopt the $Cm'cm'$ magnetic structure for simplicity. However, either of these two in-plane spin arrangements produces similar mPDF patterns, leaving our mPDF analysis unaffected.

At 5~K, neutron powder and single-crystal diffraction reveal magnetic satellites associated with $\mathbf{k}_1$ and $\mathbf{k}_2$, together with a higher-order component, 2$\mathbf{k}_1+\mathbf{k}_2$, as shown in Fig.~\ref{fig:AvgStrs}(c) and Fig.~\ref{fig:EM:TN2}(a).
Unambiguous determination of this complex magnetic structure requires high-resolution data that resolve the $\mathbf{k}_1$ and $\mathbf{k}_2$ reflections, as well as a sufficient number of magnetic reflections associated with $\mathbf{k}_1$, $\mathbf{k}_2$, and 2$\mathbf{k}_1+\mathbf{k}_2$.
Fits to the powder and single-crystal neutron data show that the magnetic superspace group $P6_{3}22.1'(0,0,\gamma_{1})t00s(0,0,\gamma_{2})h00s$ (magnetic space group number: 182.2.81.2.m180.2) best describes the data.
The comparison between calculated and measured structure factors for the nuclear reflections and 65 magnetic reflections from the single-crystal data demonstrates excellent agreement [Fig.~\ref{fig:EM:TN2}(b)].
The resulting magnetic structure is shown in Fig.~\ref{fig:EM:TN2}(c), and its individual $\mathbf{k}_1$, $\mathbf{k}_2$, and 2$\mathbf{k}_1+\mathbf{k}_2$ components are displayed in Fig.~\ref{fig:EM:TN2}(d).
The $\mathbf{k}_1$ component is a spin-density wave with moments modulated along the $c$ axis and an amplitude of 2.56(10) $\mu_{B}$, whereas the $\mathbf{k}_2$ component is an in-plane elliptical helix with amplitudes of 2.09(6)--2.25(8) $\mu_{B}$.
The 2$\mathbf{k}_1+\mathbf{k}_2$ component, arising from interference between $\mathbf{k}_1$ and $\mathbf{k}_2$, is a second elliptical helix with smaller amplitudes of 0.47(4)--0.74(3) $\mu_{B}$.
Overall, the magnetic structure of Mn$_{3}$Sn at 5~K is an amplitude-modulated conical phase with harmonic intermodulation, with moment magnitudes varying from 2.32(7) to 3.45(9) $\mu_{B}$.

To our knowledge, the magnetic structure of A-type Mn$_{3}$Sn at 5~K has not been reported previously.
Very recently, a magnetic structure at 250~K, above the $\mathbf{k}_1$-$\mathbf{k}_2$ intersection temperature but below $T_{\mathrm{N2}}$, was proposed~\cite{Chen2024}.
However, it differs from ours in two respects: (1) the $\mathbf{k_{\chi}}$ component, corresponding to  $\mathbf{k}_{2}$ in our notation, was assigned to an in-plane circular helix rather than elliptical helix, and (2) the second elliptical helix arising from the important harmonic 2$\mathbf{k}_1+\mathbf{k}_2$ component was not included in the global magnetic structure.

\begin{figure}[!ht]
    \includegraphics[width=1\linewidth]{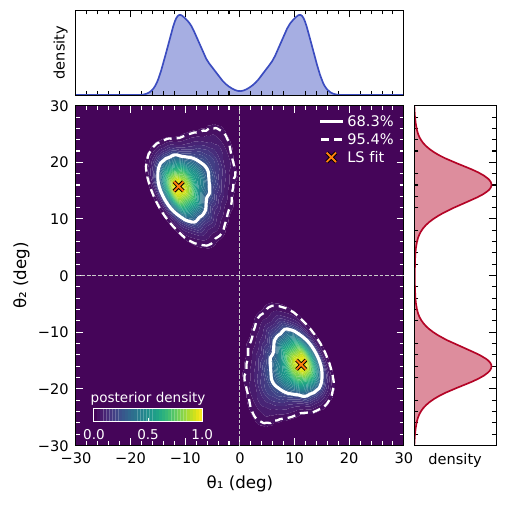}
\caption{
Joint posterior distribution of the canting angles $\theta_1$ and $\theta_2$ for the $0.5 \leq r \leq 10$~\AA{} mPDF fit.
Solid and dashed contours enclose the $68.3\%$ and $95.4\%$ credible regions, respectively.
Orange crosses indicate the best-fit values obtained from least-squares minimization.
}
\label{fig:EM:mPDF_Posterior}
\end{figure}

\textit{Field dependence of magnetization, AHE and $\rho_{xx}(T)$.}
At 300~K, while $M$--$H$ curve is linear for $H\parallel z$, magnetic hysteresis is observed with $H$ applied along the $x$ and $y$ directions [Fig.~\ref{fig:EM:MH}(a)--(c)].
Upon cooling below $T_{\mathrm{N2}}$, these magnetic hysteresis loops disappear completely [Fig.~\ref{fig:EM:MH}(a)--(c)], indicating the absence of a net ferromagnetic component in the low-T magnetic phase.
Correspondingly, the Hall resistivity becomes negligible for all three field orientations [Fig.~\ref{fig:EM:MH}(d)--(f)].
In addition, the magnetic transition at $T_{\mathrm{N2}}$ is accompanied by a distinct anomaly in the longitudinal resistivity $\rho_{xx}(T)$ for $H\parallel y$ [Fig.~\ref{fig:EM:MH}(g)], further demonstrating the coupling between the magnetic reconstruction and electronic transport.
No detectable anomaly is observed for $H\parallel x$ or $H\parallel z$, highlighting the anisotropic transport response.
 
\textit{Bayesian analysis of the out-of-plane canting.}
We quantified the canting angles $\theta_1$ and $\theta_2$ by Markov chain Monte Carlo analysis of the mPDF over $0.5\leq r\leq10.0~\text{\AA}$, using a Gaussian likelihood with uncertainties propagated from $S(Q)$ and uniform priors of $-30^\circ\leq\theta_{1,2}\leq30^\circ$~\cite{Fancher2016,Metz2018}.
Four random-walk Metropolis chains of 26{,}000 draws, each with 1000 burn-in draws, yielded 100{,}000 posterior samples, with acceptance fractions of $0.206$--$0.297$ and symmetry-reduced $\hat{R}\leq1.001$.
The posterior is bimodal under the physically equivalent global spin-reversal symmetry $(\theta_1,\theta_2)\rightarrow(-\theta_1,-\theta_2)$; adopting $\theta_1\geq0$ gives $\theta_1=9.86^{+2.67}_{-3.60}{}^\circ$ and $\theta_2=-15.97^{+3.49}_{-3.35}{}^\circ$ at 68\% credibility.
The corresponding 95\% credible intervals, $2.56^\circ\leq\theta_1\leq14.65^\circ$ and $-23.00^\circ\leq\theta_2\leq-8.25^\circ$, exclude zero and favor finite, oppositely directed out-of-plane canting.
The posterior density map and the 68.3\% and 95.4\% credible-region contours in Fig.~\ref{fig:EM:mPDF_Posterior} were obtained from a Gaussian kernel density estimate of the samples, symmetrized to populate both sign-related modes equally, with a bandwidth of $0.8$ times Scott's rule; all quoted intervals are direct sample quantiles, independent of this smoothing.
Computation details are presented in SM~\cite{SMat}.

 \end{bibunit}
 
\clearpage

\setcounter{page}{1}
\pagenumbering{arabic} 

\setcounter{figure}{0}
\renewcommand{\thefigure}{S\arabic{figure}}

\setcounter{table}{0}
\renewcommand{\thetable}{S\arabic{table}}

\setcounter{equation}{0}
\renewcommand{\theequation}{S\arabic{equation}}

\setcounter{section}{0}
\renewcommand{\thesection}{S\arabic{section}}
\begin{bibunit}
 
\onecolumngrid
\begin{center}
    {\bfseries\large Supplemental Material for \\ 
    ``Anomalous Hall Response Induced by Correlated Disorder \\in the Breathing Kagome Lattice Mn$_{3}$Sn'' \par}
    
    \vspace{1.0em}
    
    {\small 
    Tsung-Han Yang,$^{1,*}$ Seng Huat Lee,$^{2,3,*}$ Hengxin Tan,$^{4}$ Yuanpeng Zhang,$^{1}$ Benjamin A. Frandsen,$^{5}$ \\
    V\'aclav Pet\v{r}\'{i}\v{c}ek,$^{6}$ Huibo Cao,$^{1}$ Daniel Olds,$^{7}$ Matthew G. Tucker,$^{1}$ Jiaqiang Yan,$^{8}$ \\
    Binghai Yan,$^{3,\dagger}$ Zhiqiang Mao,$^{2,3,9,\ddagger}$ and Qiang Zhang$^{1,\S}$ \par}
    
    \vspace{0.8em}
    
    {\footnotesize
    $^1$\textit{Neutron Scattering Division, Oak Ridge National Laboratory, Oak Ridge, Tennessee 37831, USA} \\
    $^2$\textit{2D Crystal Consortium, Materials Research Institute, The Pennsylvania State University, University Park, Pennsylvania 16802, USA} \\
    $^3$\textit{Department of Physics, The Pennsylvania State University, University Park, Pennsylvania 16802, USA} \\
    $^4$\textit{Department of Condensed Matter Physics, Weizmann Institute of Science, Rehovot 7610001, Israel} \\
    $^5$\textit{Department of Physics and Astronomy, Brigham Young University, Provo, UT 84602, USA} \\
    $^6$\textit{Institute of Physics of the Czech Academy of Sciences, Prague, Czech Republic} \\
    $^7$\textit{National Synchrotron Light Source II, Brookhaven National Laboratory, Upton, New York 11973, USA} \\
    $^8$\textit{Materials Science and Technology Division, Oak Ridge National Laboratory, Oak Ridge, Tennessee 37831, USA} \\
    $^9$\textit{Department of Materials Science and Engineering, The Pennsylvania State University, University Park, Pennsylvania 16802, USA} \par}
    
    \vspace{0.5em}
    
    {\footnotesize 
    $^*$These authors contributed equally to this work. \\
    $^\dagger$binghai.yan@psu.edu; $^\ddagger$zim1@psu.edu; $^\S$zhangq6@ornl.gov \par}
\end{center}

\vspace{1.5em}

\twocolumngrid

\section{Sample synthesis and characterization}
Polycrystalline Mn$_3$Sn samples were synthesized from elemental Mn and Sn mixed in the stoichiometric molar ratio of 3:1.
The mixture was homogenized at $1050^\circ\mathrm{C}$ overnight and subsequently annealed at $900^\circ\mathrm{C}$ for one week.
Single crystals of Mn$_3$Sn were grown from a Sn-rich melt with a nominal Mn:Sn molar ratio of 0.7:0.3.
The mixture was first homogenized at $1000^\circ\mathrm{C}$ overnight and then slowly cooled to $900^\circ\mathrm{C}$ at a rate of $1^\circ\mathrm{C}\cdot\mathrm{h}^{-1}$.

\begin{figure}[ht!]
    \includegraphics[width=1\linewidth]{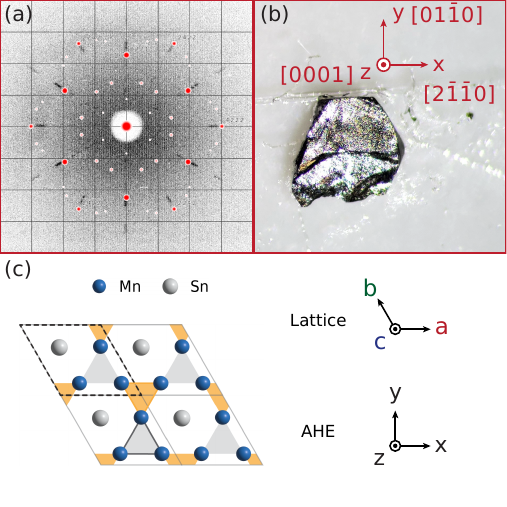}
\caption{
Single-crystal sample and crystallographic orientations of Mn$_3$Sn.
(a) Laue diffraction pattern used for orientation determination.
(b) Determined crystallographic axes.
(c) Corresponding directions in the Mn$_3$Sn unit cell.
}
\label{fig:SI:Sample}
\end{figure}

The chemical composition of the single-crystal sample was characterized by wavelength-dispersive X-ray spectroscopy (WDS) at 15 different locations across the crystal surface.
The measurements yield average Mn and Sn concentrations of $75.03(6)\pm0.02(8)$ and $24.96(4)\pm0.02(8),\mathrm{at.\%}$, respectively, corresponding to a composition of Mn$_{3.002}$Sn.
The individual measurements and compositional statistics are summarized in Table~\ref{tab:SI:WDS}.

\begin{table*}[ht!]
\caption{
WDS compositional analysis of Mn$_3$Sn measured at different locations.
The average Mn and Sn concentrations are $75.03(6)\pm0.02(8)$ and
$24.96(4)\pm0.02(8)\,\mathrm{at.\%}$, respectively, corresponding to Mn$_{3.002}$Sn.
}
\label{tab:EDS}
\centering
\setlength{\tabcolsep}{7pt}
\begin{tabular}{c ccc ccc cc}
\hline\hline
& \multicolumn{3}{c}{Weight (\%)} 
& \multicolumn{3}{c}{Atomic (\%)} 
& \multicolumn{2}{c}{Std. dev. wt.\ (\%)} \\
\cline{2-4}
\cline{5-7}
\cline{8-9}
Location & Mn & Sn & Total & Mn & Sn & Total & Mn & Sn \\
\hline
1/1 & 59.813 & 43.163 & 102.975 & 74.961 & 25.039 & 100 & 0.749 & -- \\
1/2 & 59.649 & 42.871 & 102.519 & 75.037 & 24.963 & 100 & 0.747 & -- \\
1/3 & 59.805 & 43.002 & 102.807 & 75.029 & 24.971 & 100 & 0.749 & -- \\
1/4 & 59.636 & 42.983 & 102.619 & 74.984 & 25.016 & 100 & 0.747 & -- \\
1/5 & 59.761 & 43.037 & 102.798 & 75.000 & 25.000 & 100 & 0.749 & -- \\
2/1 & 61.963 & 44.120 & 106.083 & 75.212 & 24.788 & 100 & 0.774 & -- \\
2/2 & 59.013 & 42.582 & 101.595 & 74.963 & 25.037 & 100 & 0.738 & -- \\
2/3 & 58.370 & 42.222 & 100.593 & 74.916 & 25.084 & 100 & 0.732 & -- \\
2/4 & 57.623 & 42.344 & 99.967  & 74.619 & 25.381 & 100 & 0.725 & -- \\
2/5 & 57.258 & 42.410 & 99.668  & 74.469 & 25.531 & 100 & 0.723 & -- \\
3/1 & 60.051 & 40.855 & 100.906 & 76.051 & 23.949 & 100 & 0.752 & -- \\
3/2 & 59.405 & 41.448 & 100.853 & 75.588 & 24.412 & 100 & 0.745 & -- \\
3/3 & 58.647 & 41.711 & 100.358 & 75.233 & 24.767 & 100 & 0.736 & -- \\
3/4 & 58.215 & 42.106 & 100.321 & 74.918 & 25.082 & 100 & 0.732 & -- \\
3/5 & 57.707 & 42.543 & 100.250 & 74.558 & 25.442 & 100 & 0.726 & -- \\
\hline\hline
\end{tabular}
\label{tab:SI:WDS}
\end{table*}

The crystallographic orientations of the cleaved Mn$_3$Sn single crystals were determined by back-reflection Laue X-ray diffraction, as shown in Fig.~\ref{fig:SI:Sample}.
The Laue patterns were indexed to identify the principal crystallographic axes and to verify the orientation of the exposed crystal facets [Fig.~\ref{fig:SI:Sample}(a)].
Figure~\ref{fig:SI:Sample}(c) shows the relation between the Cartesian $x$, $y$, and $z$ axes used for transport measurements and the crystallographic axes.
The oriented crystals were then used for magnetization and electrical-transport measurements with the magnetic field and current applied along the specified crystallographic directions.

\section{Single crystal x-ray diffraction}
Single-crystal X-ray diffraction measurements were performed on Mn$_{3}$Sn at 295 and 220~K using a Rigaku XtaLAB Synergy-S diffractometer with Mo K$\alpha$ radiation ($\lambda=0.71073$~\AA).
At both temperatures, the diffraction data were well described by the hexagonal $P6_3/mmc$ structure, with no evidence for a symmetry-lowering structural transition in both magnetic ordered phases.
At 295~K, the refined lattice parameters are $a=b=5.7227(4)$~\AA{} and $c=4.5684(6)$~\AA{}, whereas at 220~K they decrease to $a=b=5.6857(5)$~\AA{} and $c=4.5423(6)$~\AA{}.
The structural refinements show good agreement with the measured intensities, yielding $R_1=0.0175$ and $wR_2=0.0483$ at 295~K and $R_1=0.0240$ and $wR_2=0.0611$ at 220~K for reflections satisfying $I\geq2\sigma(I)$.
The crystallographic parameters and refinement statistics are summarized in Tables~\ref{tab:SI:XRD_TN1} and~\ref{tab:SI:XRD_TN2}.
These results confirm that the average crystallographic symmetry remains hexagonal across two magnetic transitions.
\begin{figure*}[ht!]
    \includegraphics[width=1\linewidth]{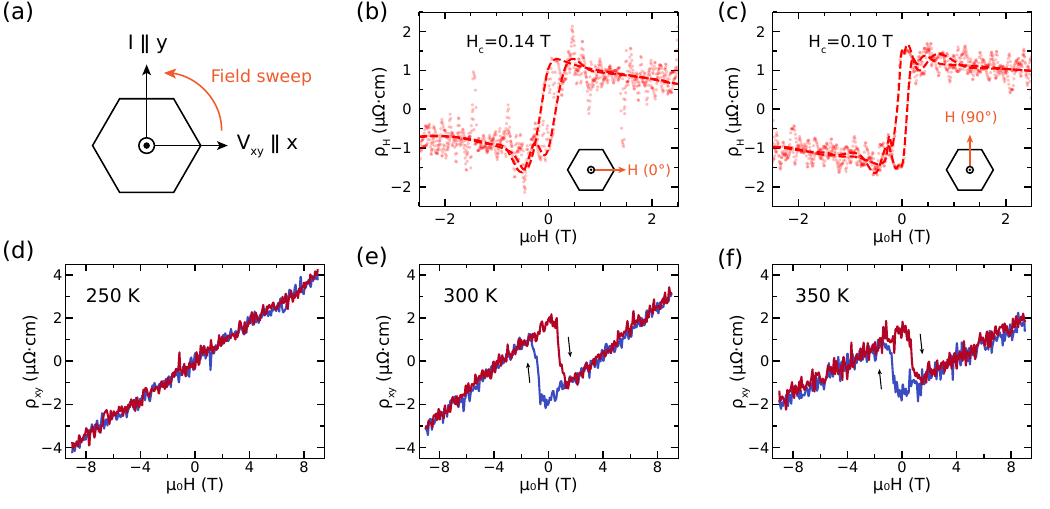}
\caption{
Hall resistivity of \MnSn{} for representative field orientations and temperatures.
(a) Measurement geometry with $I\parallel y$, $V_{xy}\parallel x$, and the magnetic field applied within the kagome plane.
(b),(c) Field-dependent Hall resistivity $\rho_H$ measured at $\theta=0^\circ$ and $90^\circ$, respectively.
For both orientations, Hall switching occurs below $H_c=0.14~\mathrm{T}$, substantially lower than the coercive field $H_c\approx0.8~\mathrm{T}$ of the out-of-plane-field Hall response.
(d)--(f) Field-dependent $\rho_{xy}$ measured at 250, 300, and 350~K, respectively, showing the temperature evolution of the anomalous Hall response.
}
\label{fig:SI_in-plane-rotation}
\end{figure*}
\begin{table*}[ht!]
\caption{Crystal data and structure refinement for \MnSn{} at 295~K.}
\begin{tabular}{ll}
\hline
Crystal system & Hexagonal \\
Space group & $P6_3/mmc$ \\
$a$ (\AA) & 5.7227(4) \\
$b$ (\AA) & 5.7227(4) \\
$c$ (\AA) & 4.5684(6) \\
$\alpha$ (°) & 90 \\
$\beta$ (°) & 90 \\
$\gamma$ (°) & 120 \\
Volume (\AA$^3$) & 129.57(2) \\
$Z$ & 2 \\
$\rho_{\text{calc}}$ (g cm$^{-3}$) & 7.267 \\
$\mu$ (mm$^{-1}$) & 23.416 \\
$F(000)$ & 250.0 \\
Crystal size (mm$^3$) & 0.251 $\times$ 0.161 $\times$ 0.052 \\
Radiation & Mo K$\alpha$ ($\lambda = 0.71073$ \AA) \\
$2\Theta$ range (°) & 8.224 to 60.656 \\
Index ranges & $-8 \leq h \leq 7$, $-8 \leq k \leq 8$, $-6 \leq l \leq 6$ \\
Reflections collected & 6478 \\
Independent reflections & 92 [$R_{\text{int}} = 0.0730$, $R_\sigma = 0.0101$] \\
Data/restraints/parameters & 92/0/9 \\
Goodness-of-fit on $F^2$ & 1.185 \\
Final $R$ indexes [$I \geq 2\sigma(I)$] & $R_1 = 0.0175$, $wR_2 = 0.0483$ \\
Final $R$ indexes [all data] & $R_1 = 0.0178$, $wR_2 = 0.0483$ \\
Largest diff. peak/hole (e \AA$^{-3}$) & 1.31 / $-0.69$ \\
\hline
\end{tabular}
\label{tab:SI:XRD_TN1}
\end{table*}

\begin{table*}[ht]
\caption{Crystal data and structure refinement for \MnSn{} at 220~K.}
\begin{tabular}{ll}
\hline
Crystal system & Hexagonal \\
Space group & $P6_3/mmc$ \\
$a$ (\AA) & 5.6857(5) \\
$b$ (\AA) & 5.6857(5) \\
$c$ (\AA) & 4.5423(6) \\
$\alpha$ (°) & 90 \\
$\beta$ (°) & 90 \\
$\gamma$ (°) & 120 \\
Volume (\AA$^3$) & 127.17(3) \\
$Z$ & 2 \\
$\rho_{\text{calc}}$ (g cm$^{-3}$) & 7.404 \\
$\mu$ (mm$^{-1}$) & 23.858 \\
$F(000)$ & 250.0 \\
Crystal size (mm$^3$) & 0.251 $\times$ 0.161 $\times$ 0.052 \\
Radiation & Mo K$\alpha$ ($\lambda = 0.71073$ \AA) \\
$2\Theta$ range (°) & 8.278 to 60.688 \\
Index ranges & $-6 \leq h \leq 7$, $-7 \leq k \leq 6$, $-6 \leq l \leq 6$ \\
Reflections collected & 856 \\
Independent reflections & 89 [$R_\text{int} = 0.0544$, $R_\sigma = 0.0184$] \\
Data/restraints/parameters & 89/0/8 \\
Goodness-of-fit on $F^2$ & 1.176 \\
Final $R$ indexes [$I \geq 2\sigma(I)$] & $R_1 = 0.0240$, $wR_2 = 0.0611$ \\
Final $R$ indexes [all data] & $R_1 = 0.0269$, $wR_2 = 0.0628$ \\
Largest diff. peak/hole (e \AA$^{-3}$) & 1.26 / $-0.68$ \\
\hline
\end{tabular}
\label{tab:SI:XRD_TN2}
\end{table*}

\section{Angle- and temperature-dependent Hall measurements}
To exclude a spurious contribution to $\sigma_{yx}$ from a small in-plane component of a misaligned magnetic field, we measured the Hall response with the field applied within the kagome plane [Fig.~\ref{fig:SI_in-plane-rotation}(a)].
For the representative orientations $\theta=0^\circ$ and $90^\circ$, the Hall response switches below $\mu_0H_{\mathrm{c}}=0.14~\mathrm{T}$ [Fig.~\ref{fig:SI_in-plane-rotation}(b),(c)].
This characteristic field scale is substantially smaller than $\mu_0H_{\mathrm{c}}\approx0.8~\mathrm{T}$ observed in the $\sigma_{yx}$ configuration, ruling out a simple field-misalignment origin of the observed $\sigma_{yx}$. In addition, the coercive fields of $\rho_{xy}$ show a temperature dependence and disappear below \(T_{\mathrm{N2}}\), further indicating that the hysteretic $\sigma_{yx}$ is an intrinsic response rather than an experimental artifact \textbf{} [Fig.~\ref{fig:SI_in-plane-rotation}(d)--(f)].

\section{Neutron diffraction experiment}
Neutron powder diffraction measurements were performed on the time-of-flight diffractometer POWGEN at the Spallation Neutron Source, Oak Ridge National Laboratory.
A cryofurnace equipped with a low-temperature insert was used to cover the temperature range from 5 to 500~K.
Temperature-dependent ramping measurements were performed at a constant rate of 2~K/min, and the cryofurnace temperature was calibrated prior to the measurements.
Two incident-neutron frames with central wavelengths of 0.8 and 2.665~\AA{} were employed, providing complementary $d$-spacing ranges of approximately 0.15--8 and 1--22~\AA{}, respectively.
Single-crystal neutron diffraction measurements were performed on DEMAND at the High Flux Isotope Reactor, Oak Ridge National Laboratory.
A closed-cycle refrigerator was used to reach temperatures down to 5~K.
Diffraction data were collected at 5~K in four-circle geometry using neutrons with a wavelength of 1.542~\AA{}, selected by a bent Si(331) monochromator~\cite{Chakoumakos2011}.
The crystal and magnetic structures were determined by refinement of the powder and single-crystal neutron diffraction data using \textsf{GSAS-II}~\cite{Toby2013} and \textsf{JANA2020}~\cite{Petricek2023}. The symmetry-allowed magnetic models are analyzed by  ISODISTORT~\cite{Campbell2006}.

\begin{figure}[t!]
    \includegraphics[width=1\linewidth]{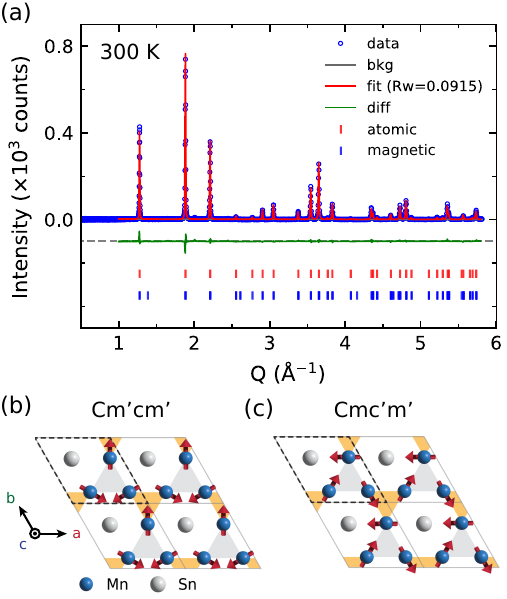}
\caption{
(a) Rietveld refinement of neutron powder-diffraction data for Mn$_3$Sn at 300 K using the inverse triangular magnetic structure with the magnetic space group $Cm'cm'$.
Open circles represent the measured intensities, the solid line is the calculated profile, and the lower curve shows the difference between observed and calculated intensities.
Tick marks indicate the positions of the nuclear and magnetic Bragg reflections.
(b) Refined magnetic structure with the magnetic space group $Cm'cm'$.
(c) Previous reported magnetic structure with $Cmc'm'$ magnetic space group~\cite{Cederholm2026}.
}
\label{fig:SI:TN1}
\end{figure}

\begin{table*}[ht!]
\caption{\label{tab:magnetic_structure}
Structural and magnetic parameters of Mn$_3$Sn at $5~\mathrm{K}$ in the
magnetic superspace group
$P6_{3}22.1^{\prime}(0,0,\gamma_{1})t00s
(0,0,\gamma_{2})h00s$.
Coordinates are given in the JANA setting.
The Fourier coefficients and moment amplitudes are in
$\mu_{\mathrm{B}}$/Mn. Parentheses denote one standard uncertainty;
values without uncertainties were constrained by symmetry.}
\centering
\begin{ruledtabular}
\begin{tabular}{lcccccc}

\multicolumn{7}{c}{Crystallographic parameters} \\
Atom & Wyckoff & Occ. & $x$ & $y$ & $z$
     & $U_{\mathrm{iso}}$ (\AA$^2$) \\
\colrule
Mn & $6h$ & 1 & 0.838420 & 0.676840 & 0.500000 & 0.0152(41) \\
Sn & $2c$ & 1 & $1/3$    & $2/3$    & $1/2$    & 0.0089(45) \\

\colrule
\multicolumn{7}{c}{Magnetic Fourier coefficients} \\
Atom & Component & Term & $M_a$ & $M_b$ & $M_c$
     & Amplitude \\
\colrule
Mn & $\mathbf{k}_1$ & sine   &  0.03(6) &  0       &  2.56(8)   & 2.56(10) \\
Mn & $\mathbf{k}_1$ & cosine &  0.10(4) &  0.20(7) &  0         & 0.17(8)  \\
Mn & $\mathbf{k}_2$ & sine   &  2.23(7) &  0       & $-0.17(8)$ & 2.23(11) \\
Mn & $\mathbf{k}_2$ & cosine &  1.20(4) &  2.40(8) &  0         & 2.08(9)  \\
Mn & $\mathbf{k}_3$ & sine   &  0.76(4) &  0       &  0.07(7)   & 0.76(8)  \\
Mn & $\mathbf{k}_3$ & cosine & $-0.27(2)$ & $-0.55(4)$ & 0      & 0.47(5)  \\

\end{tabular}
\end{ruledtabular}

\begin{flushleft}
\footnotesize
The lattice parameters are $a=b=5.6573~\text{\AA}$ and
$c=4.5207~\text{\AA}$. The propagation vectors are
$\mathbf{k}_1=(0,0,0.08778)$,
$\mathbf{k}_2=(0,0,0.10490)$, and
$\mathbf{k}_3=2\mathbf{k}_1+\mathbf{k}_2=(0,0,0.28046)$.
The atomic coordinates were fixed during the refinement, and the uniform
($\mathbf{k}=0$) magnetic moment was constrained to zero.
\end{flushleft}
\end{table*}

\section{Calculation of the Hall carrier density}
The Hall carrier density was estimated from the ordinary Hall response. The Hall resistivity was first antisymmetrized as \(\rho_{yx}(H)=[\rho_{yx}(+\mu_0H)-\rho_{yx}(-\mu_0H)]/2\). The Hall coefficient \(R_H\) was obtained from the slope of the linear field-dependent region, \(R_H=d\rho_{yx}/d(\mu_0H)\), and the corresponding Hall carrier density was estimated using \(n_H=1/(eR_H)\), where \(e\) is the elementary charge. Below \(T_{\mathrm{N2}}\), \(R_H\) was determined directly from the linear field dependence of \(\rho_{yx}(H)\). In the intermediate regime, \(T_{\mathrm{N2}}<T<T_{\mathrm{N1}}\), where an anomalous Hall contribution is present, \(R_H\) was instead extracted from the linear high-field region, where the ordinary Hall response dominates.

\section{Total Scattering and Local Structure Analysis}
Neutron total-scattering measurements were performed on POWGEN at the Spallation Neutron Source using Frame 1 with a central wavelength of 0.8~\AA.
The data extend to $Q_{\mathrm{max}}=28$~\AA$^{-1}$, providing high real-space resolution for pair distribution function (PDF) analysis.

\begin{figure}[ht!]
    \includegraphics[width=1\linewidth]{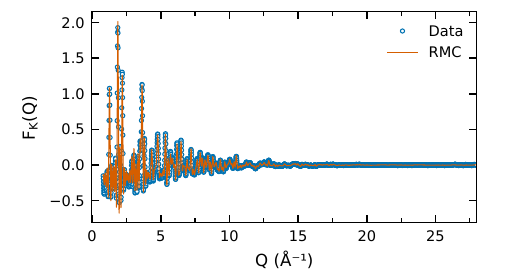}
\caption{
F$_{\mathrm{K}}(Q)$ of converged RMC fits for Mn$_3$Sn.
The corresponding G$_{\mathrm{K}}(r)$ and partial pair distribution functions are shown in the main text.
}
\label{fig:SI:RMC}
\end{figure}

Large-box reverse Monte Carlo (RMC) modeling was performed using \textsf{RMCProfile}~\cite{Tucker2007,Zhang2020}. An approximate $90\times90\times90$~\AA$^{3}$ supercell containing 46,080 atoms was constructed from the average crystallographic structure.
The atomic configurations were optimized simultaneously against the neutron total-scattering structure factor F(Q) over $0.5\leq Q\leq28$~\AA$^{-1}$, and pair distribution function G$_{K}$(r) over $2.4\leq r\leq25$~\AA.
The converged fits to $F_{\mathrm{K}}(Q)$ are shown in Fig.~\ref{fig:SI:RMC}, while the corresponding $G_{\mathrm{K}}(r)$ and partial pair distribution functions are presented in the main text.
Pairwise constraints for the Mn--Mn, Mn--Sn, and Sn--Sn partial correlations were additionally imposed to prevent unphysical atomic overlap during the modeling.
Atomic moves were accepted or rejected according to the Metropolis algorithm, and the configurations were iteratively optimized to reproduce all experimental datasets and imposed constraints simultaneously.

Following convergence, the RMC supercell was folded onto a single crystallographic unit cell to visualize the spatial distributions of the atomic positions and identify correlated local distortions.
Candidate local symmetries were examined using \textsf{FINDSYM}~\cite{Stokes2005}, which showed that the local distortion can be described by the orthorhombic $Ama2$ symmetry.
The corresponding symmetry-adapted distortion modes relative to the average hexagonal \hexsg{} structure were analyzed using \textsf{ISODISTORT}~\cite{Campbell2006}.

\begin{figure}[t!]
    \includegraphics[width=1\linewidth]{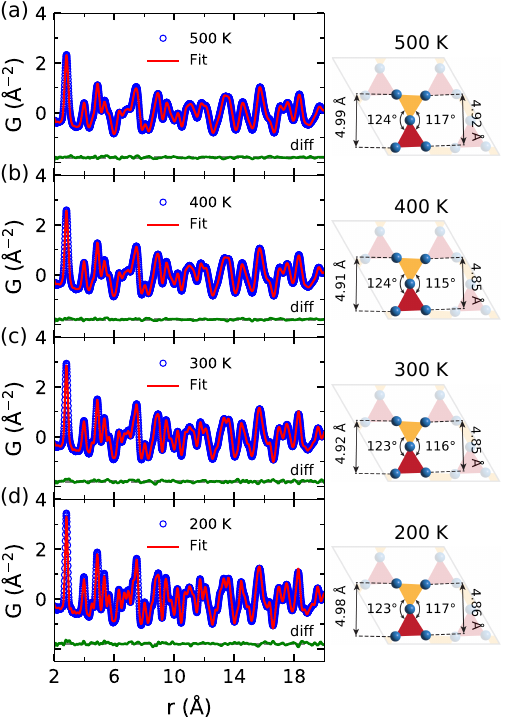}
\caption{
Temperature-dependent small-box x-ray PDF refinements of Mn$_{3}$Sn using the orthorhombic $Ama2$ model at (a) 500~K, (b) 400~K, (c) 300~K, and (d) 200~K. Refinements were performed over $2 \leq r \leq 20$~\AA{}. Insets illustrate the corresponding local Mn geometry obtained from the refinements. The $Ama2$ model describes the short-range structure across both magnetic transitions, with no evidence for an additional structural symmetry lowering at either $T_{\mathrm{N1}}$ or $T_{\mathrm{N2}}$.
}
\label{fig:SI:xPDF}
\end{figure}

Temperature-dependent synchrotron x-ray total-scattering measurements were performed at the 28-ID-1 (PDF) beamline of the National Synchrotron Light Source II, Brookhaven National Laboratory.
The polycrystalline sample was loaded into a 1~mm inner-diameter Kapton capillary, and the temperature was controlled using a nitrogen cryostream.
Measurements were carried out using an x-ray energy of 74.46~keV ($\lambda=0.1665$~\AA{}) with an amorphous-silicon PerkinElmer area detector positioned approximately 204~mm from the sample.
A Ni standard was used for detector calibration, and the background was measured using an empty Kapton capillary.
The two-dimensional scattering images were integrated using \textsf{PyFAI}~\cite{Kieffer2013} and transformed to the atomic pair distribution function $G(r)$ using \textsf{PDFgetX3}~\cite{Juhas2013}.
Temperature-dependent small-box refinements were subsequently performed over $2 \leq r \leq 20$~\AA{} using \textsf{PDFgui}~\cite{Farrow2007} and the orthorhombic $Ama2$ model obtained from the symmetry analysis of the RMC configurations.
As shown in Fig.~\ref{fig:SI:xPDF}, the $Ama2$ model consistently describes the local atomic structure at 200, 300, 400, and 500~K.
The refined local geometries show no observable evidence for further structural symmetry lowering across either magnetic transition at $T_{\mathrm{N1}}$ or $T_{\mathrm{N2}}$.

\section{Bayesian estimation of the out-of-plane canting angles}

The magnetic signal was taken as the residual of the atomic PDF refinement, $G_{\mathrm{diff}}(r)=G_{\mathrm{obs}}(r)-G^{\mathrm{nuc}}_{\mathrm{calc}}(r)$, and analyzed over $0.5\leq r\leq10$~\AA{}.
The Mn moments in the two kagome layers were parameterized by canting angles $\theta_1$ and $\theta_2$ relative to the coplanar inverse-triangular antiferromagnetic structure,
\begin{equation}
\mathbf{m}_i=\mu\left[\cos(\theta_L)\,\hat{t}_i+\sin(\theta_L)\,\hat{n}\right],
\end{equation}
where $\hat{t}_{i}$ is the fixed in-plane moment direction of site $i$, $\hat{n}$ is the kagome stacking direction, and $L\in\{1,2\}$ labels the layer containing site $i$.
The magnetic PDF was calculated following Refs.~\cite{Frandsen2014,Frandsen2015}, with the magnetic correlation length $\xi=8.85$~\AA{} and the ordered and paramagnetic scale factors, which absorb the ordered-moment magnitude, fixed at their least-squares values.
The lattice parameters and in-plane moment orientations were likewise held fixed.

The posterior distribution was evaluated using the Gaussian likelihood $\ln\mathcal{L}(\theta_1,\theta_2)=-\tfrac{1}{2}\chi^2(\theta_1,\theta_2)$, with
\begin{equation}
\chi^2(\theta_1,\theta_2)=\sum_r\frac{\left[G_{\mathrm{diff}}(r)-d_{\mathrm{mag}}(r;\theta_1,\theta_2)\right]^2}{\sigma_G^2(r)},
\end{equation}
where the sum runs over the fit range and $\sigma_G(r)$ was propagated from the experimental $S(Q)$ uncertainties through the sine Fourier transform.
Uniform priors of $-30^\circ\leq\theta_{1,2}\leq30^\circ$ were adopted, and the posterior was sampled using four independent random-walk Metropolis chains~\cite{Fancher2016,Metz2018}.
Each chain contained 26{,}000 draws, with the first 1000 discarded as burn-in, yielding 100{,}000 posterior samples with acceptance fractions of $0.206$--$0.297$.
Because every Mn site lies on a mirror plane normal to $\hat{n}$, the model is invariant under the combination of that mirror with time reversal, which maps $(\theta_1,\theta_2)\rightarrow(-\theta_1,-\theta_2)$, so the posterior is bimodal.
The reported posteriors are conditional distributions given the parameters held fixed above.
Convergence of the symmetry-reduced chains ($\theta_1\geq0$ mode) was confirmed by a Gelman--Rubin statistic $\widehat{R}\leq1.001$~\cite{GelmanRubin1992}.
The resulting posterior distributions and credible intervals are presented in the End Matter.
The density maps and credible-region contours shown there derive from a Gaussian kernel density estimate of the symmetrized samples with a bandwidth of $0.8$ times Scott's rule, while all quoted intervals are direct sample quantiles.

\section{Density functional theory}
All density-functional-theory (DFT) calculations were performed using the Vienna \textit{Ab initio} Simulation Package (\textsc{VASP})~\cite{Kresse1996CMS,Kresse1996PRB} within the projector augmented-wave (PAW) formalism~\cite{Blochl1994}.
The exchange-correlation functional was treated within the generalized gradient approximation (GGA) using the Perdew--Burke--Ernzerhof (PBE) parametrization~\cite{Perdew1996}.
The plane-wave kinetic-energy cutoff was set to 300~eV, and the Brillouin zone was sampled using an $8\times8\times10$ $\Gamma$-centered $k$-point mesh.
The experimentally determined crystal structures were used together with the corresponding non-coplanar antiferromagnetic configurations.
Spin-orbit coupling (SOC) was included in the electronic-structure calculations.
For the anomalous Hall conductivity (AHC) calculations, Wannier tight-binding Hamiltonians were constructed using \textsc{Wannier90}~\cite{Pizzi2020}, with Mn $s$ and $d$ and Sn $p$ orbitals chosen as the initial projection basis.
The intrinsic AHC was evaluated from the Wannier Hamiltonians using the Berry-curvature Kubo-formula approach described in Ref.~\cite{Zhang2017}.
 \putbib

\end{bibunit}  
 

\begin{thebibliography}{0}%
\makeatletter
\providecommand \@ifxundefined [1]{%
 \@ifx{#1\undefined}
}%
\providecommand \@ifnum [1]{%
 \ifnum #1\expandafter \@firstoftwo
 \else \expandafter \@secondoftwo
 \fi
}%
\providecommand \@ifx [1]{%
 \ifx #1\expandafter \@firstoftwo
 \else \expandafter \@secondoftwo
 \fi
}%
\providecommand \natexlab [1]{#1}%
\providecommand \enquote  [1]{``#1''}%
\providecommand \bibnamefont  [1]{#1}%
\providecommand \bibfnamefont [1]{#1}%
\providecommand \citenamefont [1]{#1}%
\providecommand \href@noop [0]{\@secondoftwo}%
\providecommand \href [0]{\begingroup \@sanitize@url \@href}%
\providecommand \@href[1]{\@@startlink{#1}\@@href}%
\providecommand \@@href[1]{\endgroup#1\@@endlink}%
\providecommand \@sanitize@url [0]{\catcode `\\12\catcode `\$12\catcode
  `\&12\catcode `\#12\catcode `\^12\catcode `\_12\catcode `\%12\relax}%
\providecommand \@@startlink[1]{}%
\providecommand \@@endlink[0]{}%
\providecommand \url  [0]{\begingroup\@sanitize@url \@url }%
\providecommand \@url [1]{\endgroup\@href {#1}{\urlprefix }}%
\providecommand \urlprefix  [0]{URL }%
\providecommand \Eprint [0]{\href }%
\providecommand \doibase [0]{https://doi.org/}%
\providecommand \selectlanguage [0]{\@gobble}%
\providecommand \bibinfo  [0]{\@secondoftwo}%
\providecommand \bibfield  [0]{\@secondoftwo}%
\providecommand \translation [1]{[#1]}%
\providecommand \BibitemOpen [0]{}%
\providecommand \bibitemStop [0]{}%
\providecommand \bibitemNoStop [0]{.\EOS\space}%
\providecommand \EOS [0]{\spacefactor3000\relax}%
\providecommand \BibitemShut  [1]{\csname bibitem#1\endcsname}%
\let\auto@bib@innerbib\@empty
\end{thebibliography}%


\begin{thebibliography}{48}%
\makeatletter
\providecommand \@ifxundefined [1]{%
 \@ifx{#1\undefined}
}%
\providecommand \@ifnum [1]{%
 \ifnum #1\expandafter \@firstoftwo
 \else \expandafter \@secondoftwo
 \fi
}%
\providecommand \@ifx [1]{%
 \ifx #1\expandafter \@firstoftwo
 \else \expandafter \@secondoftwo
 \fi
}%
\providecommand \natexlab [1]{#1}%
\providecommand \enquote  [1]{``#1''}%
\providecommand \bibnamefont  [1]{#1}%
\providecommand \bibfnamefont [1]{#1}%
\providecommand \citenamefont [1]{#1}%
\providecommand \href@noop [0]{\@secondoftwo}%
\providecommand \href [0]{\begingroup \@sanitize@url \@href}%
\providecommand \@href[1]{\@@startlink{#1}\@@href}%
\providecommand \@@href[1]{\endgroup#1\@@endlink}%
\providecommand \@sanitize@url [0]{\catcode `\\12\catcode `\$12\catcode
  `\&12\catcode `\#12\catcode `\^12\catcode `\_12\catcode `\%12\relax}%
\providecommand \@@startlink[1]{}%
\providecommand \@@endlink[0]{}%
\providecommand \url  [0]{\begingroup\@sanitize@url \@url }%
\providecommand \@url [1]{\endgroup\@href {#1}{\urlprefix }}%
\providecommand \urlprefix  [0]{URL }%
\providecommand \Eprint [0]{\href }%
\providecommand \doibase [0]{https://doi.org/}%
\providecommand \selectlanguage [0]{\@gobble}%
\providecommand \bibinfo  [0]{\@secondoftwo}%
\providecommand \bibfield  [0]{\@secondoftwo}%
\providecommand \translation [1]{[#1]}%
\providecommand \BibitemOpen [0]{}%
\providecommand \bibitemStop [0]{}%
\providecommand \bibitemNoStop [0]{.\EOS\space}%
\providecommand \EOS [0]{\spacefactor3000\relax}%
\providecommand \BibitemShut  [1]{\csname bibitem#1\endcsname}%
\let\auto@bib@innerbib\@empty
\bibitem [{\citenamefont {Schnyder}\ \emph {et~al.}(2008)\citenamefont
  {Schnyder}, \citenamefont {Ryu}, \citenamefont {Furusaki},\ and\
  \citenamefont {Ludwig}}]{Schnyder2008}%
  \BibitemOpen
  \bibfield  {author} {\bibinfo {author} {\bibfnamefont {A.~P.}\ \bibnamefont
  {Schnyder}}, \bibinfo {author} {\bibfnamefont {S.}~\bibnamefont {Ryu}},
  \bibinfo {author} {\bibfnamefont {A.}~\bibnamefont {Furusaki}},\ and\
  \bibinfo {author} {\bibfnamefont {A.~W.~W.}\ \bibnamefont {Ludwig}},\ }\href
  {https://doi.org/10.1103/PhysRevB.78.195125} {\bibfield  {journal} {\bibinfo
  {journal} {Phys. Rev. B}\ }\textbf {\bibinfo {volume} {78}},\ \bibinfo
  {pages} {195125} (\bibinfo {year} {2008})}\BibitemShut {NoStop}%
\bibitem [{\citenamefont {Hasan}\ and\ \citenamefont {Kane}(2010)}]{Hasan2010}%
  \BibitemOpen
  \bibfield  {author} {\bibinfo {author} {\bibfnamefont {M.~Z.}\ \bibnamefont
  {Hasan}}\ and\ \bibinfo {author} {\bibfnamefont {C.~L.}\ \bibnamefont
  {Kane}},\ }\href {https://doi.org/10.1103/RevModPhys.82.3045} {\bibfield
  {journal} {\bibinfo  {journal} {Rev. Mod. Phys.}\ }\textbf {\bibinfo {volume}
  {82}},\ \bibinfo {pages} {3045} (\bibinfo {year} {2010})}\BibitemShut
  {NoStop}%
\bibitem [{\citenamefont {Qi}\ and\ \citenamefont {Zhang}(2011)}]{Qi2011}%
  \BibitemOpen
  \bibfield  {author} {\bibinfo {author} {\bibfnamefont {X.-L.}\ \bibnamefont
  {Qi}}\ and\ \bibinfo {author} {\bibfnamefont {S.-C.}\ \bibnamefont {Zhang}},\
  }\href {https://doi.org/10.1103/RevModPhys.83.1057} {\bibfield  {journal}
  {\bibinfo  {journal} {Rev. Mod. Phys.}\ }\textbf {\bibinfo {volume} {83}},\
  \bibinfo {pages} {1057} (\bibinfo {year} {2011})}\BibitemShut {NoStop}%
\bibitem [{\citenamefont {Armitage}\ \emph {et~al.}(2018)\citenamefont
  {Armitage}, \citenamefont {Mele},\ and\ \citenamefont
  {Vishwanath}}]{Armitage2018}%
  \BibitemOpen
  \bibfield  {author} {\bibinfo {author} {\bibfnamefont {N.~P.}\ \bibnamefont
  {Armitage}}, \bibinfo {author} {\bibfnamefont {E.~J.}\ \bibnamefont {Mele}},\
  and\ \bibinfo {author} {\bibfnamefont {A.}~\bibnamefont {Vishwanath}},\
  }\href {https://doi.org/10.1103/RevModPhys.90.015001} {\bibfield  {journal}
  {\bibinfo  {journal} {Rev. Mod. Phys.}\ }\textbf {\bibinfo {volume} {90}},\
  \bibinfo {pages} {015001} (\bibinfo {year} {2018})}\BibitemShut {NoStop}%
\bibitem [{\citenamefont {Watanabe}\ \emph {et~al.}(2018)\citenamefont
  {Watanabe}, \citenamefont {Po},\ and\ \citenamefont
  {Vishwanath}}]{Watanabe2018}%
  \BibitemOpen
  \bibfield  {author} {\bibinfo {author} {\bibfnamefont {H.}~\bibnamefont
  {Watanabe}}, \bibinfo {author} {\bibfnamefont {H.~C.}\ \bibnamefont {Po}},\
  and\ \bibinfo {author} {\bibfnamefont {A.}~\bibnamefont {Vishwanath}},\
  }\href {https://doi.org/10.1126/sciadv.aat8685} {\bibfield  {journal}
  {\bibinfo  {journal} {Sci. Adv.}\ }\textbf {\bibinfo {volume} {4}},\ \bibinfo
  {pages} {eaat8685} (\bibinfo {year} {2018})}\BibitemShut {NoStop}%
\bibitem [{\citenamefont {Tang}\ \emph {et~al.}(2019)\citenamefont {Tang},
  \citenamefont {Po}, \citenamefont {Vishwanath},\ and\ \citenamefont
  {Wan}}]{Tang2019}%
  \BibitemOpen
  \bibfield  {author} {\bibinfo {author} {\bibfnamefont {F.}~\bibnamefont
  {Tang}}, \bibinfo {author} {\bibfnamefont {H.~C.}\ \bibnamefont {Po}},
  \bibinfo {author} {\bibfnamefont {A.}~\bibnamefont {Vishwanath}},\ and\
  \bibinfo {author} {\bibfnamefont {X.}~\bibnamefont {Wan}},\ }\href
  {https://doi.org/10.1038/s41586-019-0937-5} {\bibfield  {journal} {\bibinfo
  {journal} {Nature}\ }\textbf {\bibinfo {volume} {566}},\ \bibinfo {pages}
  {486} (\bibinfo {year} {2019})}\BibitemShut {NoStop}%
\bibitem [{\citenamefont {Bernevig}\ \emph {et~al.}(2022)\citenamefont
  {Bernevig}, \citenamefont {Felser},\ and\ \citenamefont
  {Beidenkopf}}]{Bernevig2022}%
  \BibitemOpen
  \bibfield  {author} {\bibinfo {author} {\bibfnamefont {B.~A.}\ \bibnamefont
  {Bernevig}}, \bibinfo {author} {\bibfnamefont {C.}~\bibnamefont {Felser}},\
  and\ \bibinfo {author} {\bibfnamefont {H.}~\bibnamefont {Beidenkopf}},\
  }\href {https://doi.org/10.1038/s41586-021-04105-x} {\bibfield  {journal}
  {\bibinfo  {journal} {Nature}\ }\textbf {\bibinfo {volume} {603}},\ \bibinfo
  {pages} {41} (\bibinfo {year} {2022})}\BibitemShut {NoStop}%
\bibitem [{\citenamefont {Fang}\ \emph {et~al.}(2003)\citenamefont {Fang},
  \citenamefont {Nagaosa}, \citenamefont {Takahashi}, \citenamefont {Asamitsu},
  \citenamefont {Mathieu}, \citenamefont {Ogasawara}, \citenamefont {Yamada},
  \citenamefont {Kawasaki}, \citenamefont {Tokura},\ and\ \citenamefont
  {Terakura}}]{Fang2003}%
  \BibitemOpen
  \bibfield  {author} {\bibinfo {author} {\bibfnamefont {Z.}~\bibnamefont
  {Fang}}, \bibinfo {author} {\bibfnamefont {N.}~\bibnamefont {Nagaosa}},
  \bibinfo {author} {\bibfnamefont {K.~S.}\ \bibnamefont {Takahashi}}, \bibinfo
  {author} {\bibfnamefont {A.}~\bibnamefont {Asamitsu}}, \bibinfo {author}
  {\bibfnamefont {R.}~\bibnamefont {Mathieu}}, \bibinfo {author} {\bibfnamefont
  {T.}~\bibnamefont {Ogasawara}}, \bibinfo {author} {\bibfnamefont
  {H.}~\bibnamefont {Yamada}}, \bibinfo {author} {\bibfnamefont
  {M.}~\bibnamefont {Kawasaki}}, \bibinfo {author} {\bibfnamefont
  {Y.}~\bibnamefont {Tokura}},\ and\ \bibinfo {author} {\bibfnamefont
  {K.}~\bibnamefont {Terakura}},\ }\href
  {https://doi.org/10.1126/science.1089408} {\bibfield  {journal} {\bibinfo
  {journal} {Science}\ }\textbf {\bibinfo {volume} {302}},\ \bibinfo {pages}
  {92} (\bibinfo {year} {2003})}\BibitemShut {NoStop}%
\bibitem [{\citenamefont {Haldane}(2004)}]{Haldane2004}%
  \BibitemOpen
  \bibfield  {author} {\bibinfo {author} {\bibfnamefont {F.~D.~M.}\
  \bibnamefont {Haldane}},\ }\href
  {https://doi.org/10.1103/PhysRevLett.93.206602} {\bibfield  {journal}
  {\bibinfo  {journal} {Phys. Rev. Lett.}\ }\textbf {\bibinfo {volume} {93}},\
  \bibinfo {pages} {206602} (\bibinfo {year} {2004})}\BibitemShut {NoStop}%
\bibitem [{\citenamefont {Xiao}\ \emph {et~al.}(2010)\citenamefont {Xiao},
  \citenamefont {Chang},\ and\ \citenamefont {Niu}}]{Xiao2010}%
  \BibitemOpen
  \bibfield  {author} {\bibinfo {author} {\bibfnamefont {D.}~\bibnamefont
  {Xiao}}, \bibinfo {author} {\bibfnamefont {M.-C.}\ \bibnamefont {Chang}},\
  and\ \bibinfo {author} {\bibfnamefont {Q.}~\bibnamefont {Niu}},\ }\href
  {https://doi.org/10.1103/RevModPhys.82.1959} {\bibfield  {journal} {\bibinfo
  {journal} {Rev. Mod. Phys.}\ }\textbf {\bibinfo {volume} {82}},\ \bibinfo
  {pages} {1959} (\bibinfo {year} {2010})}\BibitemShut {NoStop}%
\bibitem [{\citenamefont {Nagaosa}\ \emph {et~al.}(2010)\citenamefont
  {Nagaosa}, \citenamefont {Sinova}, \citenamefont {Onoda}, \citenamefont
  {MacDonald},\ and\ \citenamefont {Ong}}]{Nagaosa2010}%
  \BibitemOpen
  \bibfield  {author} {\bibinfo {author} {\bibfnamefont {N.}~\bibnamefont
  {Nagaosa}}, \bibinfo {author} {\bibfnamefont {J.}~\bibnamefont {Sinova}},
  \bibinfo {author} {\bibfnamefont {S.}~\bibnamefont {Onoda}}, \bibinfo
  {author} {\bibfnamefont {A.~H.}\ \bibnamefont {MacDonald}},\ and\ \bibinfo
  {author} {\bibfnamefont {N.~P.}\ \bibnamefont {Ong}},\ }\href
  {https://doi.org/10.1103/RevModPhys.82.1539} {\bibfield  {journal} {\bibinfo
  {journal} {Rev. Mod. Phys.}\ }\textbf {\bibinfo {volume} {82}},\ \bibinfo
  {pages} {1539} (\bibinfo {year} {2010})}\BibitemShut {NoStop}%
\bibitem [{\citenamefont {Ikhlas}\ \emph {et~al.}(2017)\citenamefont {Ikhlas},
  \citenamefont {Tomita}, \citenamefont {Koretsune}, \citenamefont {Suzuki},
  \citenamefont {Nishio-Hamane}, \citenamefont {Arita}, \citenamefont {Otani},\
  and\ \citenamefont {Nakatsuji}}]{Ikhlas2017}%
  \BibitemOpen
  \bibfield  {author} {\bibinfo {author} {\bibfnamefont {M.}~\bibnamefont
  {Ikhlas}}, \bibinfo {author} {\bibfnamefont {T.}~\bibnamefont {Tomita}},
  \bibinfo {author} {\bibfnamefont {T.}~\bibnamefont {Koretsune}}, \bibinfo
  {author} {\bibfnamefont {M.-T.}\ \bibnamefont {Suzuki}}, \bibinfo {author}
  {\bibfnamefont {D.}~\bibnamefont {Nishio-Hamane}}, \bibinfo {author}
  {\bibfnamefont {R.}~\bibnamefont {Arita}}, \bibinfo {author} {\bibfnamefont
  {Y.}~\bibnamefont {Otani}},\ and\ \bibinfo {author} {\bibfnamefont
  {S.}~\bibnamefont {Nakatsuji}},\ }\href {https://doi.org/10.1038/nphys4181}
  {\bibfield  {journal} {\bibinfo  {journal} {Nat. Phys.}\ }\textbf {\bibinfo
  {volume} {13}},\ \bibinfo {pages} {1085} (\bibinfo {year}
  {2017})}\BibitemShut {NoStop}%
\bibitem [{\citenamefont {Karplus}\ and\ \citenamefont
  {Luttinger}(1954)}]{Karplus1954}%
  \BibitemOpen
  \bibfield  {author} {\bibinfo {author} {\bibfnamefont {R.}~\bibnamefont
  {Karplus}}\ and\ \bibinfo {author} {\bibfnamefont {J.~M.}\ \bibnamefont
  {Luttinger}},\ }\href {https://doi.org/10.1103/PhysRev.95.1154} {\bibfield
  {journal} {\bibinfo  {journal} {Phys. Rev.}\ }\textbf {\bibinfo {volume}
  {95}},\ \bibinfo {pages} {1154} (\bibinfo {year} {1954})}\BibitemShut
  {NoStop}%
\bibitem [{\citenamefont {Onoda}\ \emph {et~al.}(2006)\citenamefont {Onoda},
  \citenamefont {Sugimoto},\ and\ \citenamefont {Nagaosa}}]{Onoda2006}%
  \BibitemOpen
  \bibfield  {author} {\bibinfo {author} {\bibfnamefont {S.}~\bibnamefont
  {Onoda}}, \bibinfo {author} {\bibfnamefont {N.}~\bibnamefont {Sugimoto}},\
  and\ \bibinfo {author} {\bibfnamefont {N.}~\bibnamefont {Nagaosa}},\ }\href
  {https://doi.org/10.1103/PhysRevLett.97.126602} {\bibfield  {journal}
  {\bibinfo  {journal} {Phys. Rev. Lett.}\ }\textbf {\bibinfo {volume} {97}},\
  \bibinfo {pages} {126602} (\bibinfo {year} {2006})}\BibitemShut {NoStop}%
\bibitem [{\citenamefont {Chen}\ \emph {et~al.}(2014)\citenamefont {Chen},
  \citenamefont {Niu},\ and\ \citenamefont {MacDonald}}]{Chen2014}%
  \BibitemOpen
  \bibfield  {author} {\bibinfo {author} {\bibfnamefont {H.}~\bibnamefont
  {Chen}}, \bibinfo {author} {\bibfnamefont {Q.}~\bibnamefont {Niu}},\ and\
  \bibinfo {author} {\bibfnamefont {A.~H.}\ \bibnamefont {MacDonald}},\ }\href
  {https://doi.org/10.1103/PhysRevLett.112.017205} {\bibfield  {journal}
  {\bibinfo  {journal} {Phys. Rev. Lett.}\ }\textbf {\bibinfo {volume} {112}},\
  \bibinfo {pages} {017205} (\bibinfo {year} {2014})}\BibitemShut {NoStop}%
\bibitem [{\citenamefont {{\v{Z}}elezn{\'y}}\ \emph {et~al.}(2014)\citenamefont
  {{\v{Z}}elezn{\'y}}, \citenamefont {Gao}, \citenamefont {V{\'y}born{\'y}},
  \citenamefont {Zemen}, \citenamefont {Ma{\v{s}}ek}, \citenamefont {Manchon},
  \citenamefont {Wunderlich}, \citenamefont {Sinova},\ and\ \citenamefont
  {Jungwirth}}]{Zelezny2014}%
  \BibitemOpen
  \bibfield  {author} {\bibinfo {author} {\bibfnamefont {J.}~\bibnamefont
  {{\v{Z}}elezn{\'y}}}, \bibinfo {author} {\bibfnamefont {H.}~\bibnamefont
  {Gao}}, \bibinfo {author} {\bibfnamefont {K.}~\bibnamefont
  {V{\'y}born{\'y}}}, \bibinfo {author} {\bibfnamefont {J.}~\bibnamefont
  {Zemen}}, \bibinfo {author} {\bibfnamefont {J.}~\bibnamefont {Ma{\v{s}}ek}},
  \bibinfo {author} {\bibfnamefont {A.}~\bibnamefont {Manchon}}, \bibinfo
  {author} {\bibfnamefont {J.}~\bibnamefont {Wunderlich}}, \bibinfo {author}
  {\bibfnamefont {J.}~\bibnamefont {Sinova}},\ and\ \bibinfo {author}
  {\bibfnamefont {T.}~\bibnamefont {Jungwirth}},\ }\href
  {https://doi.org/10.1103/PhysRevLett.113.157201} {\bibfield  {journal}
  {\bibinfo  {journal} {Phys. Rev. Lett.}\ }\textbf {\bibinfo {volume} {113}},\
  \bibinfo {pages} {157201} (\bibinfo {year} {2014})}\BibitemShut {NoStop}%
\bibitem [{\citenamefont {K{\"u}bler}\ and\ \citenamefont
  {Felser}(2014)}]{Kubler2014}%
  \BibitemOpen
  \bibfield  {author} {\bibinfo {author} {\bibfnamefont {J.}~\bibnamefont
  {K{\"u}bler}}\ and\ \bibinfo {author} {\bibfnamefont {C.}~\bibnamefont
  {Felser}},\ }\href {https://doi.org/10.1209/0295-5075/108/67001} {\bibfield
  {journal} {\bibinfo  {journal} {Europhys. Lett.}\ }\textbf {\bibinfo {volume}
  {108}},\ \bibinfo {pages} {67001} (\bibinfo {year} {2014})}\BibitemShut
  {NoStop}%
\bibitem [{\citenamefont {Li}\ \emph {et~al.}(2017)\citenamefont {Li},
  \citenamefont {Xu}, \citenamefont {Ding}, \citenamefont {Wang}, \citenamefont
  {Shen}, \citenamefont {Lu}, \citenamefont {Zhu},\ and\ \citenamefont
  {Behnia}}]{Li2017}%
  \BibitemOpen
  \bibfield  {author} {\bibinfo {author} {\bibfnamefont {X.}~\bibnamefont
  {Li}}, \bibinfo {author} {\bibfnamefont {L.}~\bibnamefont {Xu}}, \bibinfo
  {author} {\bibfnamefont {L.}~\bibnamefont {Ding}}, \bibinfo {author}
  {\bibfnamefont {J.}~\bibnamefont {Wang}}, \bibinfo {author} {\bibfnamefont
  {M.}~\bibnamefont {Shen}}, \bibinfo {author} {\bibfnamefont {X.}~\bibnamefont
  {Lu}}, \bibinfo {author} {\bibfnamefont {Z.}~\bibnamefont {Zhu}},\ and\
  \bibinfo {author} {\bibfnamefont {K.}~\bibnamefont {Behnia}},\ }\href
  {https://doi.org/10.1103/PhysRevLett.119.056601} {\bibfield  {journal}
  {\bibinfo  {journal} {Phys. Rev. Lett.}\ }\textbf {\bibinfo {volume} {119}},\
  \bibinfo {pages} {056601} (\bibinfo {year} {2017})}\BibitemShut {NoStop}%
\bibitem [{\citenamefont {Nayak}\ \emph {et~al.}(2016)\citenamefont {Nayak},
  \citenamefont {Fischer}, \citenamefont {Sun}, \citenamefont {Yan},
  \citenamefont {Karel}, \citenamefont {Komarek}, \citenamefont {Shekhar},
  \citenamefont {Kumar}, \citenamefont {Schnelle}, \citenamefont {K{\"u}bler},
  \citenamefont {Felser},\ and\ \citenamefont {Parkin}}]{Nayak2016}%
  \BibitemOpen
  \bibfield  {author} {\bibinfo {author} {\bibfnamefont {A.~K.}\ \bibnamefont
  {Nayak}}, \bibinfo {author} {\bibfnamefont {J.~E.}\ \bibnamefont {Fischer}},
  \bibinfo {author} {\bibfnamefont {Y.}~\bibnamefont {Sun}}, \bibinfo {author}
  {\bibfnamefont {B.}~\bibnamefont {Yan}}, \bibinfo {author} {\bibfnamefont
  {J.}~\bibnamefont {Karel}}, \bibinfo {author} {\bibfnamefont {A.~C.}\
  \bibnamefont {Komarek}}, \bibinfo {author} {\bibfnamefont {C.}~\bibnamefont
  {Shekhar}}, \bibinfo {author} {\bibfnamefont {N.}~\bibnamefont {Kumar}},
  \bibinfo {author} {\bibfnamefont {W.}~\bibnamefont {Schnelle}}, \bibinfo
  {author} {\bibfnamefont {J.}~\bibnamefont {K{\"u}bler}}, \bibinfo {author}
  {\bibfnamefont {C.}~\bibnamefont {Felser}},\ and\ \bibinfo {author}
  {\bibfnamefont {S.~S.~P.}\ \bibnamefont {Parkin}},\ }\href
  {https://doi.org/10.1126/sciadv.1501870} {\bibfield  {journal} {\bibinfo
  {journal} {Sci. Adv.}\ }\textbf {\bibinfo {volume} {2}},\ \bibinfo {pages}
  {e1501870} (\bibinfo {year} {2016})}\BibitemShut {NoStop}%
\bibitem [{\citenamefont {Nakatsuji}\ \emph {et~al.}(2015)\citenamefont
  {Nakatsuji}, \citenamefont {Kiyohara},\ and\ \citenamefont
  {Higo}}]{Nakatsuji2015}%
  \BibitemOpen
  \bibfield  {author} {\bibinfo {author} {\bibfnamefont {S.}~\bibnamefont
  {Nakatsuji}}, \bibinfo {author} {\bibfnamefont {N.}~\bibnamefont
  {Kiyohara}},\ and\ \bibinfo {author} {\bibfnamefont {T.}~\bibnamefont
  {Higo}},\ }\href {https://doi.org/10.1038/nature15723} {\bibfield  {journal}
  {\bibinfo  {journal} {Nature}\ }\textbf {\bibinfo {volume} {527}},\ \bibinfo
  {pages} {212} (\bibinfo {year} {2015})}\BibitemShut {NoStop}%
\bibitem [{\citenamefont {Kiyohara}\ \emph {et~al.}(2016)\citenamefont
  {Kiyohara}, \citenamefont {Tomita},\ and\ \citenamefont
  {Nakatsuji}}]{Kiyohara2016}%
  \BibitemOpen
  \bibfield  {author} {\bibinfo {author} {\bibfnamefont {N.}~\bibnamefont
  {Kiyohara}}, \bibinfo {author} {\bibfnamefont {T.}~\bibnamefont {Tomita}},\
  and\ \bibinfo {author} {\bibfnamefont {S.}~\bibnamefont {Nakatsuji}},\ }\href
  {https://doi.org/10.1103/PhysRevApplied.5.064009} {\bibfield  {journal}
  {\bibinfo  {journal} {Phys. Rev. Appl.}\ }\textbf {\bibinfo {volume} {5}},\
  \bibinfo {pages} {064009} (\bibinfo {year} {2016})}\BibitemShut {NoStop}%
\bibitem [{\citenamefont {Wuttke}\ \emph {et~al.}(2019)\citenamefont {Wuttke},
  \citenamefont {Caglieris}, \citenamefont {Sykora}, \citenamefont
  {Scaravaggi}, \citenamefont {Wolter}, \citenamefont {Manna}, \citenamefont
  {S{\"u}{\ss}}, \citenamefont {Shekhar}, \citenamefont {Felser}, \citenamefont
  {B{\"u}chner},\ and\ \citenamefont {Hess}}]{Wuttke2019}%
  \BibitemOpen
  \bibfield  {author} {\bibinfo {author} {\bibfnamefont {C.}~\bibnamefont
  {Wuttke}}, \bibinfo {author} {\bibfnamefont {F.}~\bibnamefont {Caglieris}},
  \bibinfo {author} {\bibfnamefont {S.}~\bibnamefont {Sykora}}, \bibinfo
  {author} {\bibfnamefont {F.}~\bibnamefont {Scaravaggi}}, \bibinfo {author}
  {\bibfnamefont {A.~U.~B.}\ \bibnamefont {Wolter}}, \bibinfo {author}
  {\bibfnamefont {K.}~\bibnamefont {Manna}}, \bibinfo {author} {\bibfnamefont
  {V.}~\bibnamefont {S{\"u}{\ss}}}, \bibinfo {author} {\bibfnamefont
  {C.}~\bibnamefont {Shekhar}}, \bibinfo {author} {\bibfnamefont
  {C.}~\bibnamefont {Felser}}, \bibinfo {author} {\bibfnamefont
  {B.}~\bibnamefont {B{\"u}chner}},\ and\ \bibinfo {author} {\bibfnamefont
  {C.}~\bibnamefont {Hess}},\ }\href
  {https://doi.org/10.1103/PhysRevB.100.085111} {\bibfield  {journal} {\bibinfo
   {journal} {Phys. Rev. B}\ }\textbf {\bibinfo {volume} {100}},\ \bibinfo
  {pages} {085111} (\bibinfo {year} {2019})}\BibitemShut {NoStop}%
\bibitem [{\citenamefont {Kuroda}\ \emph {et~al.}(2017)\citenamefont {Kuroda},
  \citenamefont {Tomita}, \citenamefont {Suzuki}, \citenamefont {Bareille},
  \citenamefont {Nugroho}, \citenamefont {Goswami}, \citenamefont {Ochi},
  \citenamefont {Ikhlas}, \citenamefont {Nakayama}, \citenamefont {Akebi},
  \citenamefont {Noguchi}, \citenamefont {Ishii}, \citenamefont {Inami},
  \citenamefont {Ono}, \citenamefont {Kumigashira}, \citenamefont {Varykhalov},
  \citenamefont {Muro}, \citenamefont {Koretsune}, \citenamefont {Arita},
  \citenamefont {Shin}, \citenamefont {Kondo},\ and\ \citenamefont
  {Nakatsuji}}]{Kuroda2017}%
  \BibitemOpen
  \bibfield  {author} {\bibinfo {author} {\bibfnamefont {K.}~\bibnamefont
  {Kuroda}}, \bibinfo {author} {\bibfnamefont {T.}~\bibnamefont {Tomita}},
  \bibinfo {author} {\bibfnamefont {M.-T.}\ \bibnamefont {Suzuki}}, \bibinfo
  {author} {\bibfnamefont {C.}~\bibnamefont {Bareille}}, \bibinfo {author}
  {\bibfnamefont {A.~A.}\ \bibnamefont {Nugroho}}, \bibinfo {author}
  {\bibfnamefont {P.}~\bibnamefont {Goswami}}, \bibinfo {author} {\bibfnamefont
  {M.}~\bibnamefont {Ochi}}, \bibinfo {author} {\bibfnamefont {M.}~\bibnamefont
  {Ikhlas}}, \bibinfo {author} {\bibfnamefont {M.}~\bibnamefont {Nakayama}},
  \bibinfo {author} {\bibfnamefont {S.}~\bibnamefont {Akebi}}, \bibinfo
  {author} {\bibfnamefont {R.}~\bibnamefont {Noguchi}}, \bibinfo {author}
  {\bibfnamefont {R.}~\bibnamefont {Ishii}}, \bibinfo {author} {\bibfnamefont
  {N.}~\bibnamefont {Inami}}, \bibinfo {author} {\bibfnamefont
  {K.}~\bibnamefont {Ono}}, \bibinfo {author} {\bibfnamefont {H.}~\bibnamefont
  {Kumigashira}}, \bibinfo {author} {\bibfnamefont {A.}~\bibnamefont
  {Varykhalov}}, \bibinfo {author} {\bibfnamefont {T.}~\bibnamefont {Muro}},
  \bibinfo {author} {\bibfnamefont {T.}~\bibnamefont {Koretsune}}, \bibinfo
  {author} {\bibfnamefont {R.}~\bibnamefont {Arita}}, \bibinfo {author}
  {\bibfnamefont {S.}~\bibnamefont {Shin}}, \bibinfo {author} {\bibfnamefont
  {T.}~\bibnamefont {Kondo}},\ and\ \bibinfo {author} {\bibfnamefont
  {S.}~\bibnamefont {Nakatsuji}},\ }\href {https://doi.org/10.1038/nmat4987}
  {\bibfield  {journal} {\bibinfo  {journal} {Nat. Mater.}\ }\textbf {\bibinfo
  {volume} {16}},\ \bibinfo {pages} {1090} (\bibinfo {year}
  {2017})}\BibitemShut {NoStop}%
\bibitem [{\citenamefont {Yang}\ \emph {et~al.}(2017)\citenamefont {Yang},
  \citenamefont {Sun}, \citenamefont {Zhang}, \citenamefont {Shi},
  \citenamefont {Parkin},\ and\ \citenamefont {Yan}}]{Yang2017}%
  \BibitemOpen
  \bibfield  {author} {\bibinfo {author} {\bibfnamefont {H.}~\bibnamefont
  {Yang}}, \bibinfo {author} {\bibfnamefont {Y.}~\bibnamefont {Sun}}, \bibinfo
  {author} {\bibfnamefont {Y.}~\bibnamefont {Zhang}}, \bibinfo {author}
  {\bibfnamefont {W.-J.}\ \bibnamefont {Shi}}, \bibinfo {author} {\bibfnamefont
  {S.~S.~P.}\ \bibnamefont {Parkin}},\ and\ \bibinfo {author} {\bibfnamefont
  {B.}~\bibnamefont {Yan}},\ }\href {https://doi.org/10.1088/1367-2630/aa5487}
  {\bibfield  {journal} {\bibinfo  {journal} {New J. Phys.}\ }\textbf {\bibinfo
  {volume} {19}},\ \bibinfo {pages} {015008} (\bibinfo {year}
  {2017})}\BibitemShut {NoStop}%
\bibitem [{\citenamefont {Prodan}(2011)}]{Prodan2011}%
  \BibitemOpen
  \bibfield  {author} {\bibinfo {author} {\bibfnamefont {E.}~\bibnamefont
  {Prodan}},\ }\href {https://doi.org/10.1088/1751-8113/44/11/113001}
  {\bibfield  {journal} {\bibinfo  {journal} {J. Phys. A: Math. Theor.}\
  }\textbf {\bibinfo {volume} {44}},\ \bibinfo {pages} {113001} (\bibinfo
  {year} {2011})}\BibitemShut {NoStop}%
\bibitem [{\citenamefont {Keen}\ and\ \citenamefont
  {Goodwin}(2015)}]{Keen2015}%
  \BibitemOpen
  \bibfield  {author} {\bibinfo {author} {\bibfnamefont {D.~A.}\ \bibnamefont
  {Keen}}\ and\ \bibinfo {author} {\bibfnamefont {A.~L.}\ \bibnamefont
  {Goodwin}},\ }\href {https://doi.org/10.1038/nature14453} {\bibfield
  {journal} {\bibinfo  {journal} {Nature}\ }\textbf {\bibinfo {volume} {521}},\
  \bibinfo {pages} {303} (\bibinfo {year} {2015})}\BibitemShut {NoStop}%
\bibitem [{\citenamefont {Liu}\ and\ \citenamefont {Balents}(2017)}]{Liu2017}%
  \BibitemOpen
  \bibfield  {author} {\bibinfo {author} {\bibfnamefont {J.}~\bibnamefont
  {Liu}}\ and\ \bibinfo {author} {\bibfnamefont {L.}~\bibnamefont {Balents}},\
  }\href {https://doi.org/10.1103/PhysRevLett.119.087202} {\bibfield  {journal}
  {\bibinfo  {journal} {Phys. Rev. Lett.}\ }\textbf {\bibinfo {volume} {119}},\
  \bibinfo {pages} {087202} (\bibinfo {year} {2017})}\BibitemShut {NoStop}%
\bibitem [{\citenamefont {Zhang}\ \emph {et~al.}(2017)\citenamefont {Zhang},
  \citenamefont {Sun}, \citenamefont {Yang}, \citenamefont {{\v{Z}}elezn{\'y}},
  \citenamefont {Parkin}, \citenamefont {Felser},\ and\ \citenamefont
  {Yan}}]{Zhang2017}%
  \BibitemOpen
  \bibfield  {author} {\bibinfo {author} {\bibfnamefont {Y.}~\bibnamefont
  {Zhang}}, \bibinfo {author} {\bibfnamefont {Y.}~\bibnamefont {Sun}}, \bibinfo
  {author} {\bibfnamefont {H.}~\bibnamefont {Yang}}, \bibinfo {author}
  {\bibfnamefont {J.}~\bibnamefont {{\v{Z}}elezn{\'y}}}, \bibinfo {author}
  {\bibfnamefont {S.~P.~P.}\ \bibnamefont {Parkin}}, \bibinfo {author}
  {\bibfnamefont {C.}~\bibnamefont {Felser}},\ and\ \bibinfo {author}
  {\bibfnamefont {B.}~\bibnamefont {Yan}},\ }\href
  {https://doi.org/10.1103/PhysRevB.95.075128} {\bibfield  {journal} {\bibinfo
  {journal} {Phys. Rev. B}\ }\textbf {\bibinfo {volume} {95}},\ \bibinfo
  {pages} {075128} (\bibinfo {year} {2017})}\BibitemShut {NoStop}%
\bibitem [{\citenamefont {Kr{\'e}n}\ \emph {et~al.}(1975)\citenamefont
  {Kr{\'e}n}, \citenamefont {Paitz}, \citenamefont {Zimmer},\ and\
  \citenamefont {Zsoldos}}]{Kren1975}%
  \BibitemOpen
  \bibfield  {author} {\bibinfo {author} {\bibfnamefont {E.}~\bibnamefont
  {Kr{\'e}n}}, \bibinfo {author} {\bibfnamefont {J.}~\bibnamefont {Paitz}},
  \bibinfo {author} {\bibfnamefont {G.}~\bibnamefont {Zimmer}},\ and\ \bibinfo
  {author} {\bibfnamefont {{\'E}.}~\bibnamefont {Zsoldos}},\ }\href
  {https://doi.org/10.1016/0378-4363(75)90066-2} {\bibfield  {journal}
  {\bibinfo  {journal} {Physica B+C}\ }\textbf {\bibinfo {volume} {80}},\
  \bibinfo {pages} {226} (\bibinfo {year} {1975})}\BibitemShut {NoStop}%
\bibitem [{\citenamefont {Feng}\ \emph {et~al.}(2006)\citenamefont {Feng},
  \citenamefont {Li}, \citenamefont {Ren}, \citenamefont {Li}, \citenamefont
  {Li}, \citenamefont {Li}, \citenamefont {Zhang},\ and\ \citenamefont
  {Zhang}}]{Feng2006}%
  \BibitemOpen
  \bibfield  {author} {\bibinfo {author} {\bibfnamefont {W.~J.}\ \bibnamefont
  {Feng}}, \bibinfo {author} {\bibfnamefont {D.}~\bibnamefont {Li}}, \bibinfo
  {author} {\bibfnamefont {W.~J.}\ \bibnamefont {Ren}}, \bibinfo {author}
  {\bibfnamefont {Y.~B.}\ \bibnamefont {Li}}, \bibinfo {author} {\bibfnamefont
  {W.~F.}\ \bibnamefont {Li}}, \bibinfo {author} {\bibfnamefont
  {J.}~\bibnamefont {Li}}, \bibinfo {author} {\bibfnamefont {Y.~Q.}\
  \bibnamefont {Zhang}},\ and\ \bibinfo {author} {\bibfnamefont {Z.~D.}\
  \bibnamefont {Zhang}},\ }\href {https://doi.org/10.1103/PhysRevB.73.205105}
  {\bibfield  {journal} {\bibinfo  {journal} {Phys. Rev. B}\ }\textbf {\bibinfo
  {volume} {73}},\ \bibinfo {pages} {205105} (\bibinfo {year}
  {2006})}\BibitemShut {NoStop}%
\bibitem [{\citenamefont {Park}\ \emph {et~al.}(2018)\citenamefont {Park},
  \citenamefont {Oh}, \citenamefont {Uhl{\'i}{\v{r}}ov{\'a}}, \citenamefont
  {Jackson}, \citenamefont {De{\'a}k}, \citenamefont {Szunyogh}, \citenamefont
  {Lee}, \citenamefont {Cho}, \citenamefont {Kim}, \citenamefont {Walker},
  \citenamefont {Adroja}, \citenamefont {Sechovsk{\'y}},\ and\ \citenamefont
  {Park}}]{Park2018}%
  \BibitemOpen
  \bibfield  {author} {\bibinfo {author} {\bibfnamefont {P.}~\bibnamefont
  {Park}}, \bibinfo {author} {\bibfnamefont {J.}~\bibnamefont {Oh}}, \bibinfo
  {author} {\bibfnamefont {K.}~\bibnamefont {Uhl{\'i}{\v{r}}ov{\'a}}}, \bibinfo
  {author} {\bibfnamefont {J.}~\bibnamefont {Jackson}}, \bibinfo {author}
  {\bibfnamefont {A.}~\bibnamefont {De{\'a}k}}, \bibinfo {author}
  {\bibfnamefont {L.}~\bibnamefont {Szunyogh}}, \bibinfo {author}
  {\bibfnamefont {K.~H.}\ \bibnamefont {Lee}}, \bibinfo {author} {\bibfnamefont
  {H.}~\bibnamefont {Cho}}, \bibinfo {author} {\bibfnamefont {H.-L.}\
  \bibnamefont {Kim}}, \bibinfo {author} {\bibfnamefont {H.~C.}\ \bibnamefont
  {Walker}}, \bibinfo {author} {\bibfnamefont {D.~T.}\ \bibnamefont {Adroja}},
  \bibinfo {author} {\bibfnamefont {V.}~\bibnamefont {Sechovsk{\'y}}},\ and\
  \bibinfo {author} {\bibfnamefont {J.-G.}\ \bibnamefont {Park}},\ }\href
  {https://doi.org/10.1038/s41535-018-0137-9} {\bibfield  {journal} {\bibinfo
  {journal} {npj Quantum Mater.}\ }\textbf {\bibinfo {volume} {3}},\ \bibinfo
  {pages} {63} (\bibinfo {year} {2018})}\BibitemShut {NoStop}%
\bibitem [{\citenamefont {Tomiyoshi}\ and\ \citenamefont
  {Yamaguchi}(1982)}]{Tomiyoshi1982}%
  \BibitemOpen
  \bibfield  {author} {\bibinfo {author} {\bibfnamefont {S.}~\bibnamefont
  {Tomiyoshi}}\ and\ \bibinfo {author} {\bibfnamefont {Y.}~\bibnamefont
  {Yamaguchi}},\ }\href {https://doi.org/10.1143/JPSJ.51.2478} {\bibfield
  {journal} {\bibinfo  {journal} {J. Phys. Soc. Jpn.}\ }\textbf {\bibinfo
  {volume} {51}},\ \bibinfo {pages} {2478} (\bibinfo {year}
  {1982})}\BibitemShut {NoStop}%
\bibitem [{\citenamefont {Brown}\ \emph {et~al.}(1990)\citenamefont {Brown},
  \citenamefont {Nunez}, \citenamefont {Tasset}, \citenamefont {Forsyth},\ and\
  \citenamefont {Radhakrishna}}]{Brown1990}%
  \BibitemOpen
  \bibfield  {author} {\bibinfo {author} {\bibfnamefont {P.~J.}\ \bibnamefont
  {Brown}}, \bibinfo {author} {\bibfnamefont {V.}~\bibnamefont {Nunez}},
  \bibinfo {author} {\bibfnamefont {F.}~\bibnamefont {Tasset}}, \bibinfo
  {author} {\bibfnamefont {J.~B.}\ \bibnamefont {Forsyth}},\ and\ \bibinfo
  {author} {\bibfnamefont {P.}~\bibnamefont {Radhakrishna}},\ }\href
  {https://doi.org/10.1088/0953-8984/2/47/015} {\bibfield  {journal} {\bibinfo
  {journal} {J. Phys.: Condens. Matter}\ }\textbf {\bibinfo {volume} {2}},\
  \bibinfo {pages} {9409} (\bibinfo {year} {1990})}\BibitemShut {NoStop}%
\bibitem [{\citenamefont {Chen}\ \emph {et~al.}(2024)\citenamefont {Chen},
  \citenamefont {Gaudet}, \citenamefont {Marcus}, \citenamefont {Nomoto},
  \citenamefont {Chen}, \citenamefont {Tomita}, \citenamefont {Ikhlas},
  \citenamefont {Suzuki}, \citenamefont {Zhao}, \citenamefont {Chen},
  \citenamefont {Strempfer}, \citenamefont {Arita}, \citenamefont {Nakatsuji},\
  and\ \citenamefont {Broholm}}]{Chen2024}%
  \BibitemOpen
  \bibfield  {author} {\bibinfo {author} {\bibfnamefont {Y.}~\bibnamefont
  {Chen}}, \bibinfo {author} {\bibfnamefont {J.}~\bibnamefont {Gaudet}},
  \bibinfo {author} {\bibfnamefont {G.~G.}\ \bibnamefont {Marcus}}, \bibinfo
  {author} {\bibfnamefont {T.}~\bibnamefont {Nomoto}}, \bibinfo {author}
  {\bibfnamefont {T.}~\bibnamefont {Chen}}, \bibinfo {author} {\bibfnamefont
  {T.}~\bibnamefont {Tomita}}, \bibinfo {author} {\bibfnamefont
  {M.}~\bibnamefont {Ikhlas}}, \bibinfo {author} {\bibfnamefont {H.~S.}\
  \bibnamefont {Suzuki}}, \bibinfo {author} {\bibfnamefont {Y.}~\bibnamefont
  {Zhao}}, \bibinfo {author} {\bibfnamefont {W.~C.}\ \bibnamefont {Chen}},
  \bibinfo {author} {\bibfnamefont {J.}~\bibnamefont {Strempfer}}, \bibinfo
  {author} {\bibfnamefont {R.}~\bibnamefont {Arita}}, \bibinfo {author}
  {\bibfnamefont {S.}~\bibnamefont {Nakatsuji}},\ and\ \bibinfo {author}
  {\bibfnamefont {C.}~\bibnamefont {Broholm}},\ }\href
  {https://doi.org/10.1103/PhysRevResearch.6.L032016} {\bibfield  {journal}
  {\bibinfo  {journal} {Phys. Rev. Res.}\ }\textbf {\bibinfo {volume} {6}},\
  \bibinfo {pages} {L032016} (\bibinfo {year} {2024})}\BibitemShut {NoStop}%
\bibitem [{\citenamefont {Cable}\ \emph {et~al.}(1993)\citenamefont {Cable},
  \citenamefont {Wakabayashi},\ and\ \citenamefont {Radhakrishna}}]{Cable1993}%
  \BibitemOpen
  \bibfield  {author} {\bibinfo {author} {\bibfnamefont {J.~W.}\ \bibnamefont
  {Cable}}, \bibinfo {author} {\bibfnamefont {N.}~\bibnamefont {Wakabayashi}},\
  and\ \bibinfo {author} {\bibfnamefont {P.}~\bibnamefont {Radhakrishna}},\
  }\href {https://doi.org/10.1016/0038-1098(93)90400-H} {\bibfield  {journal}
  {\bibinfo  {journal} {Solid State Commun.}\ }\textbf {\bibinfo {volume}
  {88}},\ \bibinfo {pages} {161} (\bibinfo {year} {1993})}\BibitemShut
  {NoStop}%
\bibitem [{SMa()}]{SMat}%
  \BibitemOpen
  \href@noop {} {}\bibinfo {note} {See Supplemental Material at [URL will be
  inserted by publisher] for additional experimental details, data analysis,
  and supporting results}\BibitemShut {NoStop}%
\bibitem [{\citenamefont {Li}\ \emph {et~al.}(2023)\citenamefont {Li},
  \citenamefont {Koo}, \citenamefont {Zhu}, \citenamefont {Behnia},\ and\
  \citenamefont {Yan}}]{Li2023}%
  \BibitemOpen
  \bibfield  {author} {\bibinfo {author} {\bibfnamefont {X.}~\bibnamefont
  {Li}}, \bibinfo {author} {\bibfnamefont {J.}~\bibnamefont {Koo}}, \bibinfo
  {author} {\bibfnamefont {Z.}~\bibnamefont {Zhu}}, \bibinfo {author}
  {\bibfnamefont {K.}~\bibnamefont {Behnia}},\ and\ \bibinfo {author}
  {\bibfnamefont {B.}~\bibnamefont {Yan}},\ }\href
  {https://doi.org/10.1038/s41467-023-37076-w} {\bibfield  {journal} {\bibinfo
  {journal} {Nat. Commun.}\ }\textbf {\bibinfo {volume} {14}},\ \bibinfo
  {pages} {1642} (\bibinfo {year} {2023})}\BibitemShut {NoStop}%
\bibitem [{\citenamefont {Yano}\ \emph {et~al.}(2024)\citenamefont {Yano},
  \citenamefont {Kihara}, \citenamefont {Yoneda}, \citenamefont {Vu},
  \citenamefont {Suto}, \citenamefont {Katayama}, \citenamefont {Yamaguchi},
  \citenamefont {Kuwahara}, \citenamefont {Suzuki}, \citenamefont {Saitoh},\
  and\ \citenamefont {Kashiwaya}}]{Yano2024}%
  \BibitemOpen
  \bibfield  {author} {\bibinfo {author} {\bibfnamefont {R.}~\bibnamefont
  {Yano}}, \bibinfo {author} {\bibfnamefont {S.}~\bibnamefont {Kihara}},
  \bibinfo {author} {\bibfnamefont {M.}~\bibnamefont {Yoneda}}, \bibinfo
  {author} {\bibfnamefont {H.~T.~N.}\ \bibnamefont {Vu}}, \bibinfo {author}
  {\bibfnamefont {H.}~\bibnamefont {Suto}}, \bibinfo {author} {\bibfnamefont
  {N.}~\bibnamefont {Katayama}}, \bibinfo {author} {\bibfnamefont
  {T.}~\bibnamefont {Yamaguchi}}, \bibinfo {author} {\bibfnamefont
  {M.}~\bibnamefont {Kuwahara}}, \bibinfo {author} {\bibfnamefont {M.-T.}\
  \bibnamefont {Suzuki}}, \bibinfo {author} {\bibfnamefont {K.}~\bibnamefont
  {Saitoh}},\ and\ \bibinfo {author} {\bibfnamefont {S.}~\bibnamefont
  {Kashiwaya}},\ }\href {https://doi.org/10.1063/5.0195211} {\bibfield
  {journal} {\bibinfo  {journal} {J. Chem. Phys.}\ }\textbf {\bibinfo {volume}
  {160}},\ \bibinfo {pages} {184708} (\bibinfo {year} {2024})}\BibitemShut
  {NoStop}%
\bibitem [{\citenamefont {Tucker}\ \emph {et~al.}(2007)\citenamefont {Tucker},
  \citenamefont {Keen}, \citenamefont {Dove}, \citenamefont {Goodwin},\ and\
  \citenamefont {Hui}}]{Tucker2007}%
  \BibitemOpen
  \bibfield  {author} {\bibinfo {author} {\bibfnamefont {M.~G.}\ \bibnamefont
  {Tucker}}, \bibinfo {author} {\bibfnamefont {D.~A.}\ \bibnamefont {Keen}},
  \bibinfo {author} {\bibfnamefont {M.~T.}\ \bibnamefont {Dove}}, \bibinfo
  {author} {\bibfnamefont {A.~L.}\ \bibnamefont {Goodwin}},\ and\ \bibinfo
  {author} {\bibfnamefont {Q.}~\bibnamefont {Hui}},\ }\href
  {https://doi.org/10.1088/0953-8984/19/33/335218} {\bibfield  {journal}
  {\bibinfo  {journal} {J. Phys.: Condens. Matter}\ }\textbf {\bibinfo {volume}
  {19}},\ \bibinfo {pages} {335218} (\bibinfo {year} {2007})}\BibitemShut
  {NoStop}%
\bibitem [{\citenamefont {Zhang}\ \emph {et~al.}(2020)\citenamefont {Zhang},
  \citenamefont {Eremenko}, \citenamefont {Krayzman}, \citenamefont {Tucker},\
  and\ \citenamefont {Levin}}]{Zhang2020}%
  \BibitemOpen
  \bibfield  {author} {\bibinfo {author} {\bibfnamefont {Y.}~\bibnamefont
  {Zhang}}, \bibinfo {author} {\bibfnamefont {M.}~\bibnamefont {Eremenko}},
  \bibinfo {author} {\bibfnamefont {V.}~\bibnamefont {Krayzman}}, \bibinfo
  {author} {\bibfnamefont {M.~G.}\ \bibnamefont {Tucker}},\ and\ \bibinfo
  {author} {\bibfnamefont {I.}~\bibnamefont {Levin}},\ }\href
  {https://doi.org/10.1107/S1600576720013254} {\bibfield  {journal} {\bibinfo
  {journal} {J. Appl. Crystallogr.}\ }\textbf {\bibinfo {volume} {53}},\
  \bibinfo {pages} {1509} (\bibinfo {year} {2020})}\BibitemShut {NoStop}%
\bibitem [{\citenamefont {Stokes}\ and\ \citenamefont
  {Hatch}(2005)}]{Stokes2005}%
  \BibitemOpen
  \bibfield  {author} {\bibinfo {author} {\bibfnamefont {H.~T.}\ \bibnamefont
  {Stokes}}\ and\ \bibinfo {author} {\bibfnamefont {D.~M.}\ \bibnamefont
  {Hatch}},\ }\href {https://doi.org/10.1107/S0021889804031528} {\bibfield
  {journal} {\bibinfo  {journal} {J. Appl. Crystallogr.}\ }\textbf {\bibinfo
  {volume} {38}},\ \bibinfo {pages} {237} (\bibinfo {year} {2005})}\BibitemShut
  {NoStop}%
\bibitem [{\citenamefont {Kieffer}\ and\ \citenamefont
  {Karkoulis}(2013)}]{Kieffer2013}%
  \BibitemOpen
  \bibfield  {author} {\bibinfo {author} {\bibfnamefont {J.}~\bibnamefont
  {Kieffer}}\ and\ \bibinfo {author} {\bibfnamefont {D.}~\bibnamefont
  {Karkoulis}},\ }\href {https://doi.org/10.1088/1742-6596/425/20/202012}
  {\bibfield  {journal} {\bibinfo  {journal} {J. Phys.: Conf. Ser.}\ }\textbf
  {\bibinfo {volume} {425}},\ \bibinfo {pages} {202012} (\bibinfo {year}
  {2013})}\BibitemShut {NoStop}%
\bibitem [{\citenamefont {Juh{\'a}s}\ \emph {et~al.}(2013)\citenamefont
  {Juh{\'a}s}, \citenamefont {Davis}, \citenamefont {Farrow},\ and\
  \citenamefont {Billinge}}]{Juhas2013}%
  \BibitemOpen
  \bibfield  {author} {\bibinfo {author} {\bibfnamefont {P.}~\bibnamefont
  {Juh{\'a}s}}, \bibinfo {author} {\bibfnamefont {T.}~\bibnamefont {Davis}},
  \bibinfo {author} {\bibfnamefont {C.~L.}\ \bibnamefont {Farrow}},\ and\
  \bibinfo {author} {\bibfnamefont {S.~J.~L.}\ \bibnamefont {Billinge}},\
  }\href {https://doi.org/10.1107/S0021889813005190} {\bibfield  {journal}
  {\bibinfo  {journal} {J. Appl. Crystallogr.}\ }\textbf {\bibinfo {volume}
  {46}},\ \bibinfo {pages} {560} (\bibinfo {year} {2013})}\BibitemShut
  {NoStop}%
\bibitem [{\citenamefont {Frandsen}\ \emph {et~al.}(2022)\citenamefont
  {Frandsen}, \citenamefont {Hamilton}, \citenamefont {Christensen},
  \citenamefont {Stubben},\ and\ \citenamefont {Billinge}}]{Frandsen2022}%
  \BibitemOpen
  \bibfield  {author} {\bibinfo {author} {\bibfnamefont {B.~A.}\ \bibnamefont
  {Frandsen}}, \bibinfo {author} {\bibfnamefont {P.~K.}\ \bibnamefont
  {Hamilton}}, \bibinfo {author} {\bibfnamefont {J.~A.}\ \bibnamefont
  {Christensen}}, \bibinfo {author} {\bibfnamefont {E.}~\bibnamefont
  {Stubben}},\ and\ \bibinfo {author} {\bibfnamefont {S.~J.~L.}\ \bibnamefont
  {Billinge}},\ }\href {https://doi.org/10.1107/S1600576722007257} {\bibfield
  {journal} {\bibinfo  {journal} {J. Appl. Crystallogr.}\ }\textbf {\bibinfo
  {volume} {55}},\ \bibinfo {pages} {1377} (\bibinfo {year}
  {2022})}\BibitemShut {NoStop}%
\bibitem [{\citenamefont {Song}\ \emph {et~al.}(2020)\citenamefont {Song},
  \citenamefont {Hao}, \citenamefont {Wang}, \citenamefont {Zhang},
  \citenamefont {Huang}, \citenamefont {Xing},\ and\ \citenamefont
  {Chen}}]{Song2020}%
  \BibitemOpen
  \bibfield  {author} {\bibinfo {author} {\bibfnamefont {Y.}~\bibnamefont
  {Song}}, \bibinfo {author} {\bibfnamefont {Y.}~\bibnamefont {Hao}}, \bibinfo
  {author} {\bibfnamefont {S.}~\bibnamefont {Wang}}, \bibinfo {author}
  {\bibfnamefont {J.}~\bibnamefont {Zhang}}, \bibinfo {author} {\bibfnamefont
  {Q.}~\bibnamefont {Huang}}, \bibinfo {author} {\bibfnamefont
  {X.}~\bibnamefont {Xing}},\ and\ \bibinfo {author} {\bibfnamefont
  {J.}~\bibnamefont {Chen}},\ }\href
  {https://doi.org/10.1103/PhysRevB.101.144422} {\bibfield  {journal} {\bibinfo
   {journal} {Phys. Rev. B}\ }\textbf {\bibinfo {volume} {101}},\ \bibinfo
  {pages} {144422} (\bibinfo {year} {2020})}\BibitemShut {NoStop}%
\bibitem [{\citenamefont {Cederholm}\ \emph {et~al.}(2026)\citenamefont
  {Cederholm}, \citenamefont {Xu}, \citenamefont {Guo}, \citenamefont {Ovesen},
  \citenamefont {Olsen}, \citenamefont {Krighaar}, \citenamefont {Knekna},
  \citenamefont {Soh}, \citenamefont {Lee}, \citenamefont {Qureshi},
  \citenamefont {Velamazan}, \citenamefont {Ressouche}, \citenamefont
  {Boothroyd},\ and\ \citenamefont {Jacobsen}}]{Cederholm2026}%
  \BibitemOpen
  \bibfield  {author} {\bibinfo {author} {\bibfnamefont {J.~J.}\ \bibnamefont
  {Cederholm}}, \bibinfo {author} {\bibfnamefont {Z.}~\bibnamefont {Xu}},
  \bibinfo {author} {\bibfnamefont {Y.}~\bibnamefont {Guo}}, \bibinfo {author}
  {\bibfnamefont {M.}~\bibnamefont {Ovesen}}, \bibinfo {author} {\bibfnamefont
  {T.}~\bibnamefont {Olsen}}, \bibinfo {author} {\bibfnamefont {K.~M.~L.}\
  \bibnamefont {Krighaar}}, \bibinfo {author} {\bibfnamefont {C.}~\bibnamefont
  {Knekna}}, \bibinfo {author} {\bibfnamefont {J.~R.}\ \bibnamefont {Soh}},
  \bibinfo {author} {\bibfnamefont {Y.}~\bibnamefont {Lee}}, \bibinfo {author}
  {\bibfnamefont {N.}~\bibnamefont {Qureshi}}, \bibinfo {author} {\bibfnamefont
  {J.~A.~R.}\ \bibnamefont {Velamazan}}, \bibinfo {author} {\bibfnamefont
  {E.}~\bibnamefont {Ressouche}}, \bibinfo {author} {\bibfnamefont {A.~T.}\
  \bibnamefont {Boothroyd}},\ and\ \bibinfo {author} {\bibfnamefont
  {H.}~\bibnamefont {Jacobsen}},\ }\href {https://doi.org/10.1103/dh99-3xkn}
  {\bibfield  {journal} {\bibinfo  {journal} {Phys. Rev. B}\ }\textbf {\bibinfo
  {volume} {113}},\ \bibinfo {pages} {174437} (\bibinfo {year}
  {2026})}\BibitemShut {NoStop}%
\bibitem [{\citenamefont {Fancher}\ \emph {et~al.}(2016)\citenamefont
  {Fancher}, \citenamefont {Han}, \citenamefont {Levin}, \citenamefont {Page},
  \citenamefont {Reich}, \citenamefont {Smith}, \citenamefont {Wilson},\ and\
  \citenamefont {Jones}}]{Fancher2016}%
  \BibitemOpen
  \bibfield  {author} {\bibinfo {author} {\bibfnamefont {C.~M.}\ \bibnamefont
  {Fancher}}, \bibinfo {author} {\bibfnamefont {Z.}~\bibnamefont {Han}},
  \bibinfo {author} {\bibfnamefont {I.}~\bibnamefont {Levin}}, \bibinfo
  {author} {\bibfnamefont {K.}~\bibnamefont {Page}}, \bibinfo {author}
  {\bibfnamefont {B.~J.}\ \bibnamefont {Reich}}, \bibinfo {author}
  {\bibfnamefont {R.~C.}\ \bibnamefont {Smith}}, \bibinfo {author}
  {\bibfnamefont {A.~G.}\ \bibnamefont {Wilson}},\ and\ \bibinfo {author}
  {\bibfnamefont {J.~L.}\ \bibnamefont {Jones}},\ }\href
  {https://doi.org/10.1038/srep31625} {\bibfield  {journal} {\bibinfo
  {journal} {Sci. Rep.}\ }\textbf {\bibinfo {volume} {6}},\ \bibinfo {pages}
  {31625} (\bibinfo {year} {2016})}\BibitemShut {NoStop}%
\bibitem [{\citenamefont {Metz}\ \emph {et~al.}(2018)\citenamefont {Metz},
  \citenamefont {Koch},\ and\ \citenamefont {Misture}}]{Metz2018}%
  \BibitemOpen
  \bibfield  {author} {\bibinfo {author} {\bibfnamefont {P.~C.}\ \bibnamefont
  {Metz}}, \bibinfo {author} {\bibfnamefont {R.}~\bibnamefont {Koch}},\ and\
  \bibinfo {author} {\bibfnamefont {S.~T.}\ \bibnamefont {Misture}},\ }\href
  {https://doi.org/10.1107/S1600576718011597} {\bibfield  {journal} {\bibinfo
  {journal} {J. Appl. Crystallogr.}\ }\textbf {\bibinfo {volume} {51}},\
  \bibinfo {pages} {1437} (\bibinfo {year} {2018})}\BibitemShut {NoStop}%
\end{thebibliography}%


\begin{thebibliography}{22}%
\makeatletter
\providecommand \@ifxundefined [1]{%
 \@ifx{#1\undefined}
}%
\providecommand \@ifnum [1]{%
 \ifnum #1\expandafter \@firstoftwo
 \else \expandafter \@secondoftwo
 \fi
}%
\providecommand \@ifx [1]{%
 \ifx #1\expandafter \@firstoftwo
 \else \expandafter \@secondoftwo
 \fi
}%
\providecommand \natexlab [1]{#1}%
\providecommand \enquote  [1]{``#1''}%
\providecommand \bibnamefont  [1]{#1}%
\providecommand \bibfnamefont [1]{#1}%
\providecommand \citenamefont [1]{#1}%
\providecommand \href@noop [0]{\@secondoftwo}%
\providecommand \href [0]{\begingroup \@sanitize@url \@href}%
\providecommand \@href[1]{\@@startlink{#1}\@@href}%
\providecommand \@@href[1]{\endgroup#1\@@endlink}%
\providecommand \@sanitize@url [0]{\catcode `\\12\catcode `\$12\catcode
  `\&12\catcode `\#12\catcode `\^12\catcode `\_12\catcode `\%12\relax}%
\providecommand \@@startlink[1]{}%
\providecommand \@@endlink[0]{}%
\providecommand \url  [0]{\begingroup\@sanitize@url \@url }%
\providecommand \@url [1]{\endgroup\@href {#1}{\urlprefix }}%
\providecommand \urlprefix  [0]{URL }%
\providecommand \Eprint [0]{\href }%
\providecommand \doibase [0]{https://doi.org/}%
\providecommand \selectlanguage [0]{\@gobble}%
\providecommand \bibinfo  [0]{\@secondoftwo}%
\providecommand \bibfield  [0]{\@secondoftwo}%
\providecommand \translation [1]{[#1]}%
\providecommand \BibitemOpen [0]{}%
\providecommand \bibitemStop [0]{}%
\providecommand \bibitemNoStop [0]{.\EOS\space}%
\providecommand \EOS [0]{\spacefactor3000\relax}%
\providecommand \BibitemShut  [1]{\csname bibitem#1\endcsname}%
\let\auto@bib@innerbib\@empty
\bibitem [{\citenamefont {Chakoumakos}\ \emph {et~al.}(2011)\citenamefont
  {Chakoumakos}, \citenamefont {Cao}, \citenamefont {Ye}, \citenamefont
  {Stoica}, \citenamefont {Popovici}, \citenamefont {Sundaram}, \citenamefont
  {Zhou}, \citenamefont {Hicks}, \citenamefont {Lynn},\ and\ \citenamefont
  {Riedel}}]{Chakoumakos2011}%
  \BibitemOpen
  \bibfield  {author} {\bibinfo {author} {\bibfnamefont {B.~C.}\ \bibnamefont
  {Chakoumakos}}, \bibinfo {author} {\bibfnamefont {H.}~\bibnamefont {Cao}},
  \bibinfo {author} {\bibfnamefont {F.}~\bibnamefont {Ye}}, \bibinfo {author}
  {\bibfnamefont {A.~D.}\ \bibnamefont {Stoica}}, \bibinfo {author}
  {\bibfnamefont {M.}~\bibnamefont {Popovici}}, \bibinfo {author}
  {\bibfnamefont {M.}~\bibnamefont {Sundaram}}, \bibinfo {author}
  {\bibfnamefont {W.}~\bibnamefont {Zhou}}, \bibinfo {author} {\bibfnamefont
  {J.~S.}\ \bibnamefont {Hicks}}, \bibinfo {author} {\bibfnamefont {G.~W.}\
  \bibnamefont {Lynn}},\ and\ \bibinfo {author} {\bibfnamefont {R.~A.}\
  \bibnamefont {Riedel}},\ }\href {https://doi.org/10.1107/S0021889811012301}
  {\bibfield  {journal} {\bibinfo  {journal} {J. Appl. Crystallogr.}\ }\textbf
  {\bibinfo {volume} {44}},\ \bibinfo {pages} {655} (\bibinfo {year}
  {2011})}\BibitemShut {NoStop}%
\bibitem [{\citenamefont {Toby}\ and\ \citenamefont
  {Von~Dreele}(2013)}]{Toby2013}%
  \BibitemOpen
  \bibfield  {author} {\bibinfo {author} {\bibfnamefont {B.~H.}\ \bibnamefont
  {Toby}}\ and\ \bibinfo {author} {\bibfnamefont {R.~B.}\ \bibnamefont
  {Von~Dreele}},\ }\href {https://doi.org/10.1107/S0021889813003531} {\bibfield
   {journal} {\bibinfo  {journal} {J. Appl. Crystallogr.}\ }\textbf {\bibinfo
  {volume} {46}},\ \bibinfo {pages} {544} (\bibinfo {year} {2013})}\BibitemShut
  {NoStop}%
\bibitem [{\citenamefont {Pet{\v{r}}{\'i}{\v{c}}ek}\ \emph
  {et~al.}(2023)\citenamefont {Pet{\v{r}}{\'i}{\v{c}}ek}, \citenamefont
  {Palatinus}, \citenamefont {Pl{\'a}{\v{s}}il},\ and\ \citenamefont
  {Du{\v{s}}ek}}]{Petricek2023}%
  \BibitemOpen
  \bibfield  {author} {\bibinfo {author} {\bibfnamefont {V.}~\bibnamefont
  {Pet{\v{r}}{\'i}{\v{c}}ek}}, \bibinfo {author} {\bibfnamefont
  {L.}~\bibnamefont {Palatinus}}, \bibinfo {author} {\bibfnamefont
  {J.}~\bibnamefont {Pl{\'a}{\v{s}}il}},\ and\ \bibinfo {author} {\bibfnamefont
  {M.}~\bibnamefont {Du{\v{s}}ek}},\ }\href
  {https://doi.org/10.1515/zkri-2023-0005} {\bibfield  {journal} {\bibinfo
  {journal} {Z. Kristallogr. - Cryst. Mater.}\ }\textbf {\bibinfo {volume}
  {238}},\ \bibinfo {pages} {271} (\bibinfo {year} {2023})}\BibitemShut
  {NoStop}%
\bibitem [{\citenamefont {Campbell}\ \emph {et~al.}(2006)\citenamefont
  {Campbell}, \citenamefont {Stokes}, \citenamefont {Tanner},\ and\
  \citenamefont {Hatch}}]{Campbell2006}%
  \BibitemOpen
  \bibfield  {author} {\bibinfo {author} {\bibfnamefont {B.~J.}\ \bibnamefont
  {Campbell}}, \bibinfo {author} {\bibfnamefont {H.~T.}\ \bibnamefont
  {Stokes}}, \bibinfo {author} {\bibfnamefont {D.~E.}\ \bibnamefont {Tanner}},\
  and\ \bibinfo {author} {\bibfnamefont {D.~M.}\ \bibnamefont {Hatch}},\ }\href
  {https://doi.org/10.1107/S0021889806014075} {\bibfield  {journal} {\bibinfo
  {journal} {J. Appl. Crystallogr.}\ }\textbf {\bibinfo {volume} {39}},\
  \bibinfo {pages} {607} (\bibinfo {year} {2006})}\BibitemShut {NoStop}%
\bibitem [{\citenamefont {Cederholm}\ \emph {et~al.}(2026)\citenamefont
  {Cederholm}, \citenamefont {Xu}, \citenamefont {Guo}, \citenamefont {Ovesen},
  \citenamefont {Olsen}, \citenamefont {Krighaar}, \citenamefont {Knekna},
  \citenamefont {Soh}, \citenamefont {Lee}, \citenamefont {Qureshi},
  \citenamefont {Velamazan}, \citenamefont {Ressouche}, \citenamefont
  {Boothroyd},\ and\ \citenamefont {Jacobsen}}]{Cederholm2026}%
  \BibitemOpen
  \bibfield  {author} {\bibinfo {author} {\bibfnamefont {J.~J.}\ \bibnamefont
  {Cederholm}}, \bibinfo {author} {\bibfnamefont {Z.}~\bibnamefont {Xu}},
  \bibinfo {author} {\bibfnamefont {Y.}~\bibnamefont {Guo}}, \bibinfo {author}
  {\bibfnamefont {M.}~\bibnamefont {Ovesen}}, \bibinfo {author} {\bibfnamefont
  {T.}~\bibnamefont {Olsen}}, \bibinfo {author} {\bibfnamefont {K.~M.~L.}\
  \bibnamefont {Krighaar}}, \bibinfo {author} {\bibfnamefont {C.}~\bibnamefont
  {Knekna}}, \bibinfo {author} {\bibfnamefont {J.~R.}\ \bibnamefont {Soh}},
  \bibinfo {author} {\bibfnamefont {Y.}~\bibnamefont {Lee}}, \bibinfo {author}
  {\bibfnamefont {N.}~\bibnamefont {Qureshi}}, \bibinfo {author} {\bibfnamefont
  {J.~A.~R.}\ \bibnamefont {Velamazan}}, \bibinfo {author} {\bibfnamefont
  {E.}~\bibnamefont {Ressouche}}, \bibinfo {author} {\bibfnamefont {A.~T.}\
  \bibnamefont {Boothroyd}},\ and\ \bibinfo {author} {\bibfnamefont
  {H.}~\bibnamefont {Jacobsen}},\ }\href {https://doi.org/10.1103/dh99-3xkn}
  {\bibfield  {journal} {\bibinfo  {journal} {Phys. Rev. B}\ }\textbf {\bibinfo
  {volume} {113}},\ \bibinfo {pages} {174437} (\bibinfo {year}
  {2026})}\BibitemShut {NoStop}%
\bibitem [{\citenamefont {Tucker}\ \emph {et~al.}(2007)\citenamefont {Tucker},
  \citenamefont {Keen}, \citenamefont {Dove}, \citenamefont {Goodwin},\ and\
  \citenamefont {Hui}}]{Tucker2007}%
  \BibitemOpen
  \bibfield  {author} {\bibinfo {author} {\bibfnamefont {M.~G.}\ \bibnamefont
  {Tucker}}, \bibinfo {author} {\bibfnamefont {D.~A.}\ \bibnamefont {Keen}},
  \bibinfo {author} {\bibfnamefont {M.~T.}\ \bibnamefont {Dove}}, \bibinfo
  {author} {\bibfnamefont {A.~L.}\ \bibnamefont {Goodwin}},\ and\ \bibinfo
  {author} {\bibfnamefont {Q.}~\bibnamefont {Hui}},\ }\href
  {https://doi.org/10.1088/0953-8984/19/33/335218} {\bibfield  {journal}
  {\bibinfo  {journal} {J. Phys.: Condens. Matter}\ }\textbf {\bibinfo {volume}
  {19}},\ \bibinfo {pages} {335218} (\bibinfo {year} {2007})}\BibitemShut
  {NoStop}%
\bibitem [{\citenamefont {Zhang}\ \emph {et~al.}(2020)\citenamefont {Zhang},
  \citenamefont {Eremenko}, \citenamefont {Krayzman}, \citenamefont {Tucker},\
  and\ \citenamefont {Levin}}]{Zhang2020}%
  \BibitemOpen
  \bibfield  {author} {\bibinfo {author} {\bibfnamefont {Y.}~\bibnamefont
  {Zhang}}, \bibinfo {author} {\bibfnamefont {M.}~\bibnamefont {Eremenko}},
  \bibinfo {author} {\bibfnamefont {V.}~\bibnamefont {Krayzman}}, \bibinfo
  {author} {\bibfnamefont {M.~G.}\ \bibnamefont {Tucker}},\ and\ \bibinfo
  {author} {\bibfnamefont {I.}~\bibnamefont {Levin}},\ }\href
  {https://doi.org/10.1107/S1600576720013254} {\bibfield  {journal} {\bibinfo
  {journal} {J. Appl. Crystallogr.}\ }\textbf {\bibinfo {volume} {53}},\
  \bibinfo {pages} {1509} (\bibinfo {year} {2020})}\BibitemShut {NoStop}%
\bibitem [{\citenamefont {Stokes}\ and\ \citenamefont
  {Hatch}(2005)}]{Stokes2005}%
  \BibitemOpen
  \bibfield  {author} {\bibinfo {author} {\bibfnamefont {H.~T.}\ \bibnamefont
  {Stokes}}\ and\ \bibinfo {author} {\bibfnamefont {D.~M.}\ \bibnamefont
  {Hatch}},\ }\href {https://doi.org/10.1107/S0021889804031528} {\bibfield
  {journal} {\bibinfo  {journal} {J. Appl. Crystallogr.}\ }\textbf {\bibinfo
  {volume} {38}},\ \bibinfo {pages} {237} (\bibinfo {year} {2005})}\BibitemShut
  {NoStop}%
\bibitem [{\citenamefont {Kieffer}\ and\ \citenamefont
  {Karkoulis}(2013)}]{Kieffer2013}%
  \BibitemOpen
  \bibfield  {author} {\bibinfo {author} {\bibfnamefont {J.}~\bibnamefont
  {Kieffer}}\ and\ \bibinfo {author} {\bibfnamefont {D.}~\bibnamefont
  {Karkoulis}},\ }\href {https://doi.org/10.1088/1742-6596/425/20/202012}
  {\bibfield  {journal} {\bibinfo  {journal} {J. Phys.: Conf. Ser.}\ }\textbf
  {\bibinfo {volume} {425}},\ \bibinfo {pages} {202012} (\bibinfo {year}
  {2013})}\BibitemShut {NoStop}%
\bibitem [{\citenamefont {Juh{\'a}s}\ \emph {et~al.}(2013)\citenamefont
  {Juh{\'a}s}, \citenamefont {Davis}, \citenamefont {Farrow},\ and\
  \citenamefont {Billinge}}]{Juhas2013}%
  \BibitemOpen
  \bibfield  {author} {\bibinfo {author} {\bibfnamefont {P.}~\bibnamefont
  {Juh{\'a}s}}, \bibinfo {author} {\bibfnamefont {T.}~\bibnamefont {Davis}},
  \bibinfo {author} {\bibfnamefont {C.~L.}\ \bibnamefont {Farrow}},\ and\
  \bibinfo {author} {\bibfnamefont {S.~J.~L.}\ \bibnamefont {Billinge}},\
  }\href {https://doi.org/10.1107/S0021889813005190} {\bibfield  {journal}
  {\bibinfo  {journal} {J. Appl. Crystallogr.}\ }\textbf {\bibinfo {volume}
  {46}},\ \bibinfo {pages} {560} (\bibinfo {year} {2013})}\BibitemShut
  {NoStop}%
\bibitem [{\citenamefont {Farrow}\ \emph {et~al.}(2007)\citenamefont {Farrow},
  \citenamefont {Juh{\'a}s}, \citenamefont {Liu}, \citenamefont {Bryndin},
  \citenamefont {Bo{\v{z}}in}, \citenamefont {Bloch}, \citenamefont {Proffen},\
  and\ \citenamefont {Billinge}}]{Farrow2007}%
  \BibitemOpen
  \bibfield  {author} {\bibinfo {author} {\bibfnamefont {C.~L.}\ \bibnamefont
  {Farrow}}, \bibinfo {author} {\bibfnamefont {P.}~\bibnamefont {Juh{\'a}s}},
  \bibinfo {author} {\bibfnamefont {J.~W.}\ \bibnamefont {Liu}}, \bibinfo
  {author} {\bibfnamefont {D.}~\bibnamefont {Bryndin}}, \bibinfo {author}
  {\bibfnamefont {E.~S.}\ \bibnamefont {Bo{\v{z}}in}}, \bibinfo {author}
  {\bibfnamefont {J.}~\bibnamefont {Bloch}}, \bibinfo {author} {\bibfnamefont
  {T.}~\bibnamefont {Proffen}},\ and\ \bibinfo {author} {\bibfnamefont
  {S.~J.~L.}\ \bibnamefont {Billinge}},\ }\href
  {https://doi.org/10.1088/0953-8984/19/33/335219} {\bibfield  {journal}
  {\bibinfo  {journal} {J. Phys.: Condens. Matter}\ }\textbf {\bibinfo {volume}
  {19}},\ \bibinfo {pages} {335219} (\bibinfo {year} {2007})}\BibitemShut
  {NoStop}%
\bibitem [{\citenamefont {Frandsen}\ \emph {et~al.}(2014)\citenamefont
  {Frandsen}, \citenamefont {Yang},\ and\ \citenamefont
  {Billinge}}]{Frandsen2014}%
  \BibitemOpen
  \bibfield  {author} {\bibinfo {author} {\bibfnamefont {B.~A.}\ \bibnamefont
  {Frandsen}}, \bibinfo {author} {\bibfnamefont {X.}~\bibnamefont {Yang}},\
  and\ \bibinfo {author} {\bibfnamefont {S.~J.~L.}\ \bibnamefont {Billinge}},\
  }\href {https://doi.org/10.1107/S2053273313033081} {\bibfield  {journal}
  {\bibinfo  {journal} {Acta Crystallogr. Sect. A Found. Adv.}\ }\textbf
  {\bibinfo {volume} {70}},\ \bibinfo {pages} {3} (\bibinfo {year}
  {2014})}\BibitemShut {NoStop}%
\bibitem [{\citenamefont {Frandsen}\ and\ \citenamefont
  {Billinge}(2015)}]{Frandsen2015}%
  \BibitemOpen
  \bibfield  {author} {\bibinfo {author} {\bibfnamefont {B.~A.}\ \bibnamefont
  {Frandsen}}\ and\ \bibinfo {author} {\bibfnamefont {S.~J.~L.}\ \bibnamefont
  {Billinge}},\ }\href {https://doi.org/10.1107/S205327331500306X} {\bibfield
  {journal} {\bibinfo  {journal} {Acta Crystallogr. Sect. A Found. Adv.}\
  }\textbf {\bibinfo {volume} {71}},\ \bibinfo {pages} {325} (\bibinfo {year}
  {2015})}\BibitemShut {NoStop}%
\bibitem [{\citenamefont {Fancher}\ \emph {et~al.}(2016)\citenamefont
  {Fancher}, \citenamefont {Han}, \citenamefont {Levin}, \citenamefont {Page},
  \citenamefont {Reich}, \citenamefont {Smith}, \citenamefont {Wilson},\ and\
  \citenamefont {Jones}}]{Fancher2016}%
  \BibitemOpen
  \bibfield  {author} {\bibinfo {author} {\bibfnamefont {C.~M.}\ \bibnamefont
  {Fancher}}, \bibinfo {author} {\bibfnamefont {Z.}~\bibnamefont {Han}},
  \bibinfo {author} {\bibfnamefont {I.}~\bibnamefont {Levin}}, \bibinfo
  {author} {\bibfnamefont {K.}~\bibnamefont {Page}}, \bibinfo {author}
  {\bibfnamefont {B.~J.}\ \bibnamefont {Reich}}, \bibinfo {author}
  {\bibfnamefont {R.~C.}\ \bibnamefont {Smith}}, \bibinfo {author}
  {\bibfnamefont {A.~G.}\ \bibnamefont {Wilson}},\ and\ \bibinfo {author}
  {\bibfnamefont {J.~L.}\ \bibnamefont {Jones}},\ }\href
  {https://doi.org/10.1038/srep31625} {\bibfield  {journal} {\bibinfo
  {journal} {Sci. Rep.}\ }\textbf {\bibinfo {volume} {6}},\ \bibinfo {pages}
  {31625} (\bibinfo {year} {2016})}\BibitemShut {NoStop}%
\bibitem [{\citenamefont {Metz}\ \emph {et~al.}(2018)\citenamefont {Metz},
  \citenamefont {Koch},\ and\ \citenamefont {Misture}}]{Metz2018}%
  \BibitemOpen
  \bibfield  {author} {\bibinfo {author} {\bibfnamefont {P.~C.}\ \bibnamefont
  {Metz}}, \bibinfo {author} {\bibfnamefont {R.}~\bibnamefont {Koch}},\ and\
  \bibinfo {author} {\bibfnamefont {S.~T.}\ \bibnamefont {Misture}},\ }\href
  {https://doi.org/10.1107/S1600576718011597} {\bibfield  {journal} {\bibinfo
  {journal} {J. Appl. Crystallogr.}\ }\textbf {\bibinfo {volume} {51}},\
  \bibinfo {pages} {1437} (\bibinfo {year} {2018})}\BibitemShut {NoStop}%
\bibitem [{\citenamefont {Gelman}\ and\ \citenamefont
  {Rubin}(1992)}]{GelmanRubin1992}%
  \BibitemOpen
  \bibfield  {author} {\bibinfo {author} {\bibfnamefont {A.}~\bibnamefont
  {Gelman}}\ and\ \bibinfo {author} {\bibfnamefont {D.~B.}\ \bibnamefont
  {Rubin}},\ }\href {https://doi.org/10.1214/ss/1177011136} {\bibfield
  {journal} {\bibinfo  {journal} {Stat. Sci.}\ }\textbf {\bibinfo {volume}
  {7}},\ \bibinfo {pages} {457} (\bibinfo {year} {1992})}\BibitemShut {NoStop}%
\bibitem [{\citenamefont {Kresse}\ and\ \citenamefont
  {Furthm{\"u}ller}(1996{\natexlab{a}})}]{Kresse1996CMS}%
  \BibitemOpen
  \bibfield  {author} {\bibinfo {author} {\bibfnamefont {G.}~\bibnamefont
  {Kresse}}\ and\ \bibinfo {author} {\bibfnamefont {J.}~\bibnamefont
  {Furthm{\"u}ller}},\ }\href {https://doi.org/10.1016/0927-0256(96)00008-0}
  {\bibfield  {journal} {\bibinfo  {journal} {Comput. Mater. Sci.}\ }\textbf
  {\bibinfo {volume} {6}},\ \bibinfo {pages} {15} (\bibinfo {year}
  {1996}{\natexlab{a}})}\BibitemShut {NoStop}%
\bibitem [{\citenamefont {Kresse}\ and\ \citenamefont
  {Furthm{\"u}ller}(1996{\natexlab{b}})}]{Kresse1996PRB}%
  \BibitemOpen
  \bibfield  {author} {\bibinfo {author} {\bibfnamefont {G.}~\bibnamefont
  {Kresse}}\ and\ \bibinfo {author} {\bibfnamefont {J.}~\bibnamefont
  {Furthm{\"u}ller}},\ }\href {https://doi.org/10.1103/PhysRevB.54.11169}
  {\bibfield  {journal} {\bibinfo  {journal} {Phys. Rev. B}\ }\textbf {\bibinfo
  {volume} {54}},\ \bibinfo {pages} {11169} (\bibinfo {year}
  {1996}{\natexlab{b}})}\BibitemShut {NoStop}%
\bibitem [{\citenamefont {Bl{\"o}chl}(1994)}]{Blochl1994}%
  \BibitemOpen
  \bibfield  {author} {\bibinfo {author} {\bibfnamefont {P.~E.}\ \bibnamefont
  {Bl{\"o}chl}},\ }\href {https://doi.org/10.1103/PhysRevB.50.17953} {\bibfield
   {journal} {\bibinfo  {journal} {Phys. Rev. B}\ }\textbf {\bibinfo {volume}
  {50}},\ \bibinfo {pages} {17953} (\bibinfo {year} {1994})}\BibitemShut
  {NoStop}%
\bibitem [{\citenamefont {Perdew}\ \emph {et~al.}(1996)\citenamefont {Perdew},
  \citenamefont {Burke},\ and\ \citenamefont {Ernzerhof}}]{Perdew1996}%
  \BibitemOpen
  \bibfield  {author} {\bibinfo {author} {\bibfnamefont {J.~P.}\ \bibnamefont
  {Perdew}}, \bibinfo {author} {\bibfnamefont {K.}~\bibnamefont {Burke}},\ and\
  \bibinfo {author} {\bibfnamefont {M.}~\bibnamefont {Ernzerhof}},\ }\href
  {https://doi.org/10.1103/PhysRevLett.77.3865} {\bibfield  {journal} {\bibinfo
   {journal} {Phys. Rev. Lett.}\ }\textbf {\bibinfo {volume} {77}},\ \bibinfo
  {pages} {3865} (\bibinfo {year} {1996})}\BibitemShut {NoStop}%
\bibitem [{\citenamefont {Pizzi}\ \emph {et~al.}(2020)\citenamefont {Pizzi},
  \citenamefont {Vitale}, \citenamefont {Arita}, \citenamefont {Bl{\"u}gel},
  \citenamefont {Freimuth}, \citenamefont {G{\'e}ranton}, \citenamefont
  {Gibertini}, \citenamefont {Gresch}, \citenamefont {Johnson}, \citenamefont
  {Koretsune}, \citenamefont {Iba{\~n}ez-Azpiroz}, \citenamefont {Lee},
  \citenamefont {Lihm}, \citenamefont {Marchand}, \citenamefont {Marrazzo},
  \citenamefont {Mokrousov}, \citenamefont {Mustafa}, \citenamefont {Nohara},
  \citenamefont {Nomura}, \citenamefont {Paulatto}, \citenamefont {Ponc{\'e}},
  \citenamefont {Ponweiser}, \citenamefont {Qiao}, \citenamefont {Th{\"o}le},
  \citenamefont {Tsirkin}, \citenamefont {Wierzbowska}, \citenamefont
  {Marzari}, \citenamefont {Vanderbilt}, \citenamefont {Souza}, \citenamefont
  {Mostofi},\ and\ \citenamefont {Yates}}]{Pizzi2020}%
  \BibitemOpen
  \bibfield  {author} {\bibinfo {author} {\bibfnamefont {G.}~\bibnamefont
  {Pizzi}}, \bibinfo {author} {\bibfnamefont {V.}~\bibnamefont {Vitale}},
  \bibinfo {author} {\bibfnamefont {R.}~\bibnamefont {Arita}}, \bibinfo
  {author} {\bibfnamefont {S.}~\bibnamefont {Bl{\"u}gel}}, \bibinfo {author}
  {\bibfnamefont {F.}~\bibnamefont {Freimuth}}, \bibinfo {author}
  {\bibfnamefont {G.}~\bibnamefont {G{\'e}ranton}}, \bibinfo {author}
  {\bibfnamefont {M.}~\bibnamefont {Gibertini}}, \bibinfo {author}
  {\bibfnamefont {D.}~\bibnamefont {Gresch}}, \bibinfo {author} {\bibfnamefont
  {C.}~\bibnamefont {Johnson}}, \bibinfo {author} {\bibfnamefont
  {T.}~\bibnamefont {Koretsune}}, \bibinfo {author} {\bibfnamefont
  {J.}~\bibnamefont {Iba{\~n}ez-Azpiroz}}, \bibinfo {author} {\bibfnamefont
  {H.}~\bibnamefont {Lee}}, \bibinfo {author} {\bibfnamefont {J.~M.}\
  \bibnamefont {Lihm}}, \bibinfo {author} {\bibfnamefont {D.}~\bibnamefont
  {Marchand}}, \bibinfo {author} {\bibfnamefont {A.}~\bibnamefont {Marrazzo}},
  \bibinfo {author} {\bibfnamefont {Y.}~\bibnamefont {Mokrousov}}, \bibinfo
  {author} {\bibfnamefont {J.~I.}\ \bibnamefont {Mustafa}}, \bibinfo {author}
  {\bibfnamefont {Y.}~\bibnamefont {Nohara}}, \bibinfo {author} {\bibfnamefont
  {Y.}~\bibnamefont {Nomura}}, \bibinfo {author} {\bibfnamefont
  {L.}~\bibnamefont {Paulatto}}, \bibinfo {author} {\bibfnamefont
  {S.}~\bibnamefont {Ponc{\'e}}}, \bibinfo {author} {\bibfnamefont
  {T.}~\bibnamefont {Ponweiser}}, \bibinfo {author} {\bibfnamefont
  {J.}~\bibnamefont {Qiao}}, \bibinfo {author} {\bibfnamefont {F.}~\bibnamefont
  {Th{\"o}le}}, \bibinfo {author} {\bibfnamefont {S.~S.}\ \bibnamefont
  {Tsirkin}}, \bibinfo {author} {\bibfnamefont {M.}~\bibnamefont
  {Wierzbowska}}, \bibinfo {author} {\bibfnamefont {N.}~\bibnamefont
  {Marzari}}, \bibinfo {author} {\bibfnamefont {D.}~\bibnamefont {Vanderbilt}},
  \bibinfo {author} {\bibfnamefont {I.}~\bibnamefont {Souza}}, \bibinfo
  {author} {\bibfnamefont {A.~A.}\ \bibnamefont {Mostofi}},\ and\ \bibinfo
  {author} {\bibfnamefont {J.~R.}\ \bibnamefont {Yates}},\ }\href
  {https://doi.org/10.1088/1361-648X/ab51ff} {\bibfield  {journal} {\bibinfo
  {journal} {J. Phys.: Condens. Matter}\ }\textbf {\bibinfo {volume} {32}},\
  \bibinfo {pages} {165902} (\bibinfo {year} {2020})}\BibitemShut {NoStop}%
\bibitem [{\citenamefont {Zhang}\ \emph {et~al.}(2017)\citenamefont {Zhang},
  \citenamefont {Sun}, \citenamefont {Yang}, \citenamefont {{\v{Z}}elezn{\'y}},
  \citenamefont {Parkin}, \citenamefont {Felser},\ and\ \citenamefont
  {Yan}}]{Zhang2017}%
  \BibitemOpen
  \bibfield  {author} {\bibinfo {author} {\bibfnamefont {Y.}~\bibnamefont
  {Zhang}}, \bibinfo {author} {\bibfnamefont {Y.}~\bibnamefont {Sun}}, \bibinfo
  {author} {\bibfnamefont {H.}~\bibnamefont {Yang}}, \bibinfo {author}
  {\bibfnamefont {J.}~\bibnamefont {{\v{Z}}elezn{\'y}}}, \bibinfo {author}
  {\bibfnamefont {S.~P.~P.}\ \bibnamefont {Parkin}}, \bibinfo {author}
  {\bibfnamefont {C.}~\bibnamefont {Felser}},\ and\ \bibinfo {author}
  {\bibfnamefont {B.}~\bibnamefont {Yan}},\ }\href
  {https://doi.org/10.1103/PhysRevB.95.075128} {\bibfield  {journal} {\bibinfo
  {journal} {Phys. Rev. B}\ }\textbf {\bibinfo {volume} {95}},\ \bibinfo
  {pages} {075128} (\bibinfo {year} {2017})}\BibitemShut {NoStop}%
\end{thebibliography}%
\end{document}